\documentclass[amsmat,amssymb,amsfonts,aps,prX,twocolumn,showpacs,superscriptaddress]{revtex4-1}
\usepackage{graphicx}
\usepackage{dcolumn}
\usepackage{bm}
\usepackage{amsmath}

\newcommand{\beq}{\begin{equation}}  
\newcommand{\eeq}{\end{equation}}  
\newcommand{\beqa}{\begin{eqnarray}}  
\newcommand{\eeqa}{\end{eqnarray}}

\usepackage{xcolor}

\newcommand{\RKKY}{Ruderman_Kittel_pr_1954, Kasuda_pthephys_1956,Yosida_pr_1957}

\newcommand{\spinchains}{Crommie_Lutz_science_1993,Hirjibehedin_Lutz_Science_2006,Wahl_Simon_PRL_2007,
Bode_Heide_nature_2007,Loth_Baumann_science_2012,Nadj_Drozdov_science_2014,
Spinelli_Bryant_naturematerials_2014,Choi_Gupta_jp_2014,Ruby_Pientka_prl_2015,Bryant_Toskovic_nanol_2015,
Choi_Robles_arXiv_2015,Yan_Malavolti_arxiv_2016}

\begin{document}

\title{RKKY oscillations in the spin relaxation rates of atomic scale nanomagnets}

\author{F. Delgado}
\affiliation{Centro de F\'{i}sica de Materiales, CSIC-UPV/EHU,\\
Donostia International Physics Center (DIPC),\\
 Paseo Manuel de Lardizabal 4-5, E-20018 Donostia-San Sebasti\'an, Spain}
\affiliation{IKERBASQUE, Basque Foundation for Science, E-48013 Bilbao, Spain}

\author{J. Fern\'andez-Rossier}
\affiliation{QuantaLab, International Iberian Nanotechnology Laboratory (INL),
Av. Mestre Jos\'e Veiga, 4715-310 Braga, Portugal}
\affiliation{Departamento de F\'{i}sica Aplicada, Universidad de Alicante,  Spain}


\begin{abstract}
Exchange interactions  with itinerant electrons are known to act as a relaxation mechanism for individual local  spins.
The same exchange interactions are also known 
to induce the so called RKKY indirect exchange interaction between two otherwise decoupled local spins.
   Here we show that both the spin relaxation and the RKKY coupling can be seen as the dissipative and reactive response to the coupling of the local spins with the itinerant electrons. We thereby  predict that  the spin relaxation rates of   magnetic nanostructures of exchanged coupled local spins, such as  as nanoengineered spin chains, have  an oscillatory dependence on $k_F d$ , where $k_F$ is the Fermi wavenumber and $d$ is the inter-spin distance,  very much like the celebrated oscillations in the RKKY  interaction.     
  We demonstrate that both $T_1$ and $T_2$ can be enhanced or suppressed, compared to the single spin limit, depending on the interplay between the Fermi surface and the nanostructure geometrical arrangement.  
   Our results open  a route 
   to  engineer   spin relaxation and decoherence in atomically designed spin structures.
\end{abstract}

 
\maketitle


\section{Introduction}

The relaxation  of localized spins plays a central role in several branches of science and technology. For instance, magnetic resonance imaging is mostly based on the sensitivity of proton  spin relaxation to its environment~\cite{Brown_Cheng_book_2014},  while the sensitivity of single spin nanomagnetometers based on the localized spins of NV centers in diamond is limited by spin relaxation time~\cite{Jelezko_Wrachtrup_physstsol_2006,Luan_Grinolds_srep_2015,Balasubramanian_Chan_nature_2008}.  Analogously, the upper  time limit  for quantum computations based on spin-qubits 
is determined by the spin decoherence time of these systems.   Therefore,  there is  an enormous interest in understanding and engineering spin relaxation in multi-spin structures, where there is a competition between the internal  spin-spin interactions  in the system of interest and the spin interactions with its  environment. 

For more than 6 decades now,  it has been known that spin-exchange interaction between local spins and itinerant electrons in a conductor  results both in the spin relaxation of the local spins, as proposed by Korringa~\cite{Korringa_physica_1950},   as well as in an effective RKKY spin-spin exchange~\cite{\RKKY}.    Initially proposed for  nuclear spins  hyperfine-coupled to the conduction electrons,  these two physical concepts  were  applied right away to electronic local moments interacting via Kondo-like exchange with the conduction electrons.   Both the  Korringa spin relaxation  rate and the RKKY interactions are proportional to $(\rho J)^2$ where $\rho$ is the density of states of the conduction electrons at the Fermi energy and $J$ is the magnitude of the Kondo-like exchange~\cite{Kittel_Fong_book_1987}. As we discuss below, the Korringa spin relaxation and the RKKY interaction can actually be understood as the dissipative and reactive forces induced by the coupling to the conduction electrons.  
Interestingly,  the Korringa spin relaxation has most often  been studied as a single localized-spin phenomenon whereas the RKKY interaction is clearly a multi-spin concept.    Here we study  the Korringa spin relaxation of  chains of  localized spins with lattice parameter  $d$ and we find that their spin relaxation rates also have an oscillating dependence on $k_F d$, which opens the door for engineering the spin relaxation.   

Our work is motivated in part by the striking progress in scanning tunnelling microscopy (STM) techniques that permits probing magnetic nanostructures with atomic precision, such as molecular magnets~\cite{Kahle_Deng_nanolett_2011,Burgess_Malavolti_2015} and atomically engineered spin chains deposited on conducting surfaces~\cite{\spinchains}.  The fabrication of atomic scale spin chains, either by 
self-assembly~\cite{Bode_Heide_nature_2007,Nadj_Drozdov_science_2014,Ruby_Pientka_prl_2015} or placing atoms one by one using STM~\cite{Crommie_Lutz_science_1993,Hirjibehedin_Lutz_Science_2006,Loth_Baumann_science_2012,
Spinelli_Bryant_naturematerials_2014,Choi_Gupta_jp_2014,Bryant_Toskovic_nanol_2015,
Choi_Robles_arXiv_2015,Yan_Malavolti_arxiv_2016},  is now becoming routine.  Moreover, STM also provides a route to probe the magnetization~\cite{Loth_Baumann_science_2012,Spinelli_Bryant_naturematerials_2014},  the spin excitations~\cite{Hirjibehedin_Lutz_Science_2006,Loth_Baumann_science_2012,
 Spinelli_Bryant_naturematerials_2014,Bryant_Toskovic_nanol_2015,Choi_Robles_arXiv_2015} or  the spin relaxation  dynamics~\cite{Loth_Baumann_science_2012,Spinelli_Bryant_naturematerials_2014,Yan_Malavolti_arxiv_2016}  of atomically engineered spin chains.  And last but not least,  
the decoherence time  $T_2$  of an individual magnetic adatom has been recently measured by electron spin resonance~\cite{Baumann_Paul_science_2015}, opening the way to use individual spins, probed with STM, as magnetometers~\cite{natterer2016}. 
    
In many instances~\cite{Hirjibehedin_Lutz_Science_2006,Loth_Bergmann_natphys_2010,Loth_Baumann_science_2012,
 Spinelli_Bryant_naturematerials_2014,Choi_Gupta_jp_2014,Jacobson_Herden_natcom_2015}, the magnetic atoms and the underlying metallic surface are separated by a decoupling layer,  
which reduces the strength of the Kondo exchange interactions and slows down the spin relaxation.  
  In that situation, the spin excitations and the spin dynamics of the spin array have been successfully modelled treating  the Kondo interactions of the atomic spins with the conduction electrons perturbatively~\cite{Delgado_Rossier_prb_2010,Gauyacq_Lorente_pss_2012,Delgado_Rossier_prl_2012,Ternes_njp_2015},  and at the same time, treating the  Heisenberg and anisotropy terms exactly by means of numerical diagonalization.  
  
  The purpose of this work is to analyze the  problem of spin relaxation of finite spin chains computed to second order in the  Kondo coupling with an electron gas. We will  pay special attention to the interplay between the Fermi wavelength $k_F$ and the inter-spin distance $d$ that arises from the phase coherence of the scattering wave functions at different atoms in a given chain. 
  Whereas the single spin Kondo system is one of the most studied problems in condensed matter physics~\cite{Hewson_book_1997},  the analogous problem with many spins has received comparatively less attention, yet  there is a substantial body of work~\cite{Jones_Varma_PRL_1988,Zachar96,Tsunetsugu_Sigrist_rmp_1997,Prokof_Stamp_jlt_1996,Neto_Jones_PhysRevB_2000,Shah_Millis_PRL_2003,Misra_book_2007,Lobos_Cazalilla_PRB_2012}.

The rest of this manuscript is organized as follows. In Sec. II we review the model Hamiltonian and discuss the physical origin of the phase term in the Kondo coupling.  In Sec. III we  present the main results of the dissipative dynamics of the chain when the Kondo coupling is treated perturbatively. The interference effects in the Kondo coupling with a spin dimer are analyzed in Sec.~\ref{secdimer}, while the influence on the spin wave relaxation of ferromagnetic chains are studied in Sec.~\ref{SW_ferro}. The effects on decoherence of spin chains are analyzed in Sec.~\ref{T2_deco}. Finally, the main conclusions are summarized in Sec.~\ref{discuss} together with a detailed discussion of the possible consequences. Some important technical details are  discussed  in the Appendix.

\begin{figure}[hbt]
\includegraphics[width=1.\linewidth]{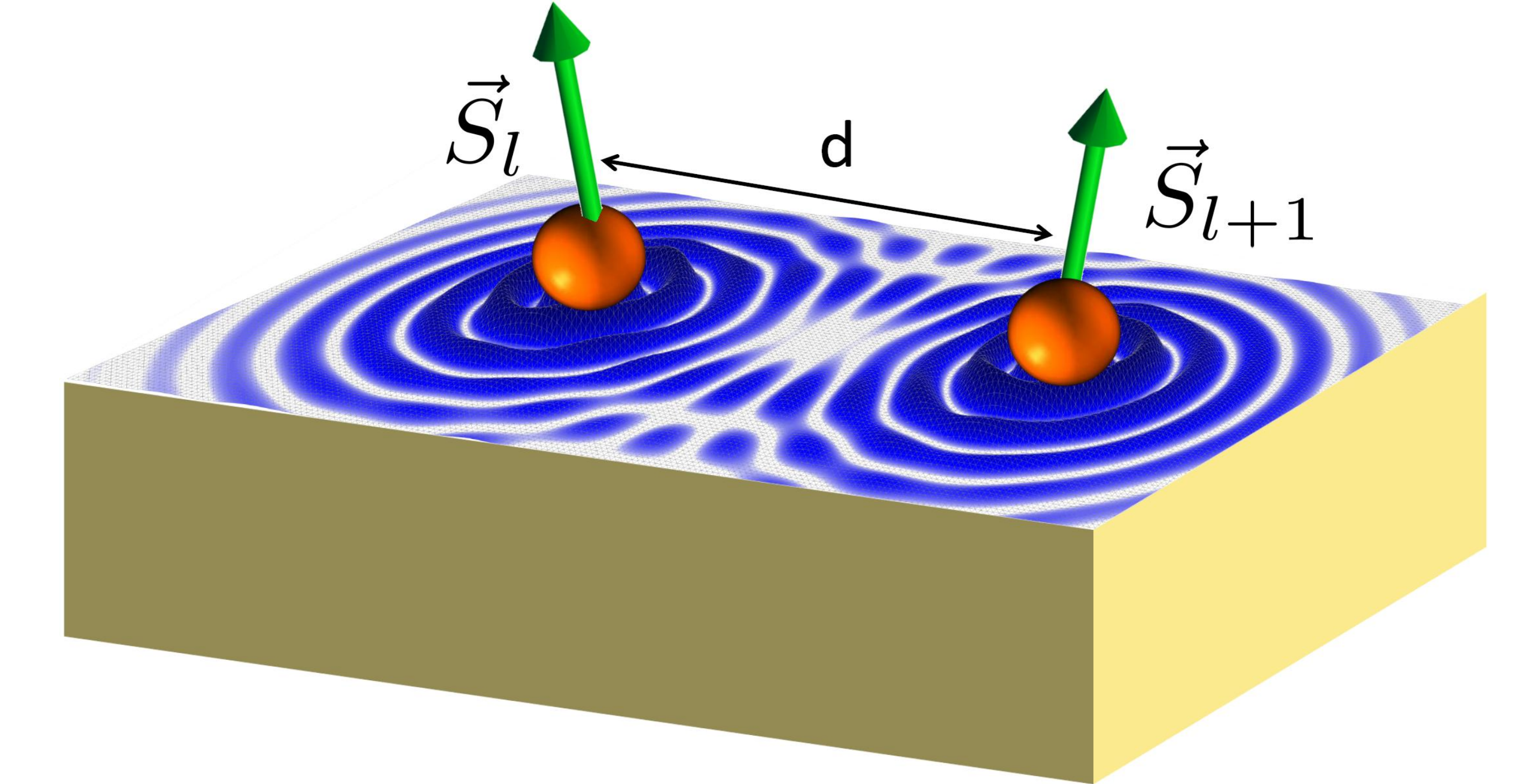}
\caption{ \label{fig1}(Color online). 
Scheme of the scattering of conduction electrons, forming interfering waves due to their interaction  with two localized
surface spins $\vec S_l$ and $\vec S_{l+1}$ separated by a distance $d$. The  interference pattern  depends on the distance between atoms $d$ and affects their spin relaxation.}
\end{figure}

\section{The Kondo-coupled spin chain Hamiltonian}
The system of interest is a magnetic nanostructure, such as  spin chain with  $N$ atoms (tipically $N<50$) ~\cite{Crommie_Lutz_science_1993,Hirjibehedin_Lutz_Science_2006,Loth_Baumann_science_2012,
Spinelli_Bryant_naturematerials_2014,Choi_Gupta_jp_2014,Bryant_Toskovic_nanol_2015,
Choi_Robles_arXiv_2015,Yan_Malavolti_arxiv_2016} or stacks of magnetic molecules~\cite{Chen_Fu_prl_2008},  weakly coupled to the itinerant electrons of a nearby nonmagnetic metal. In the following we discuss the case
of  linear chains of atomic spins  Kondo coupled to a Fermi gas, but the theory can be readily applied to other geometries.  The Hamiltonian describing the whole system is the sum of 3 terms: 

\beqa
H_T={\cal H}_{\rm R}+ {\cal H}_{\rm chain} + {\cal V},
\label{hamilT}
\eeqa
 where 
 ${\cal H}_{\rm R}=\sum_{\lambda,\sigma} \epsilon_{\lambda,\sigma}
c^{\dagger}_{\lambda,\sigma}c_{\lambda,\sigma}$
 describes the free electrons at the reservoir characterized by spin $\sigma$ and other quantum numbers, such as momentum and band,  encoded in $\lambda$. The Hamiltonian of the spin array
 is denoted by    ${\cal H}_{\rm chain}$  and the Kondo  coupling between the spins in the chain and the itinerant electrons is given by  ${\cal V}$, which will be detailed below.

\subsection{The anisotropic Heisenberg  Hamiltonian} 

For simplicity, we consider  a chain of $N$ quantized spins, with first neighbor Heisenberg exchange $J_H$ and  single ion anisotropy:
\beqa
{\cal H}_{\rm chain}=
\sum_{l=1}^N {\cal H}_0(l) + J_H \sum_{l=1}^{N-1}\vec{S}_l\cdot\vec{S}_{l+1},
\label{hamilC}
\eeqa
where
\beqa
 {\cal H}_0(l)=DS_z^2(l)+E\left(S_x^2(l)-S_y^2(l)\right)
\label{hani}
\eeqa  
describes the single ion anisotropy with the lowest possible symmetry, apt for transition metals on the Cu$_2$N/Cu(100)~\cite{Hirjibehedin_Lutz_Science_2006,Loth_Baumann_science_2012,
 Spinelli_Bryant_naturematerials_2014,Loth_Bergmann_natphys_2010,Bryant_Spinelli_prl_2013,
 Bryant_Toskovic_nanol_2015}
  surface and several other surfaces such as Al$_2$O$_3$~\cite{Heinrich_Gupta_science_2004},
MgO~\cite{Rau_Baumann_scince2014},  or h-BN~\cite{Jacobson_Herden_natcom_2015}.
    We ignore second neighbor exchange Heisenberg interactions, that  probably plays a role~\cite{Rossier_prl_2009} as well as Dzyaloshinskii-Moriya interactions,  definitely permitted due to the inherent lack of inversion symmetry of surfaces~\cite{Crepieux_Lacroix_JMMM_1998}.

     It must be pointed out that once the atomic spins are Kondo coupled to the electron gas,  effective 
     long-range indirect exchange interactions emerge, which would contribute to $J_H$ in Eq. (\ref{hamilC}) as well as to longer range couplings~\cite{Khajetoorians_Steinbrecher_natcomm_2016}. These are the celebrated RKKY interactions~\cite{\RKKY},
     but also higher order super-exchange like terms~\cite{Zubi_Bihlmayer_pss_2011,Ong_Jones_epl_2011}. 
 Since we deal with the weak coupling regime, we ignore these effects altogether. 
We stress that in this regime, the most important results of this work  do not depend on the symmetry properties of the single ion Hamiltonian (\ref{hani}), nor on the nature of the exchange interactions.

The experimental systems of interest have $N$ in the range of 12 or less~\cite{Loth_Baumann_science_2012,
 Spinelli_Bryant_naturematerials_2014,Yan_Malavolti_arxiv_2016}, and $1/2\le S \lesssim  5/2$.  This permits us to treat the chain Hamiltonian exactly by numerical diagonalization. 
Depending on the values of $J_H$, $D$, $E$, $S$ and $N$, this Hamiltonian can describe a very large variety of different ground states. For instance, in the case of $D=E=0$,  we have a pure Heisenberg model,  which happens to be a  good approximation for the description of Mn spin chains on Cu$_2$N~\cite{Hirjibehedin_Lutz_Science_2006,Rossier_prl_2009}.
By contrast, chains of magnetic adatoms with large easy axis anisotropy ($-D\gg k_BT, E$) can behave as an $S=1/2$-XXZ Heisenberg model~\cite{Toskovic_berg_natphys_2016}.

\subsection{The surface-Kondo coupling \label{KdondoS}}
The Kondo coupling between the atoms in the magnetic nanostructure and the itinerant electrons takes the general form~\cite{Tsunetsugu_Sigrist_rmp_1997,Prokof_Stamp_jlt_1996,Neto_Jones_PhysRevB_2000,Shah_Millis_PRL_2003,Misra_book_2007}: 
\beqa
{\cal V}_K=  \sum_l^N  {\cal J}(l) \vec S(l) \cdot \vec s(\vec r_l),
\label{VKONDO}
\eeqa
where $\vec S(l)$ is the spin of the $l$-magnetic adatom and $\vec s(\vec r_l)$ is the surface spin density evaluated at the position $\vec r_l$ of the $l$ magnetic atom:
\beqa
\vec s(\vec r_l)=\sum_{\vec k\vec k'\sigma\sigma'}e^{i\left(\vec k-\vec k'\right)\cdot \vec r_l}\frac{\vec \sigma_{\sigma\sigma'}}{2}
c^{\dagger}_{\vec k,\sigma} c_{\vec k'\sigma'}.
\label{Sdensity}
\eeqa
Here  $\vec{\sigma}$ denotes the Pauli matrices vector. In the following we assume that  the strength of the Kondo interaction, governed by ${\cal J}(l)$ is the same for all the atoms in the chain, i.e.,  ${\cal J}(l)=J$.   This is expected to be the case in experiments of chains of magnetic adatoms adsorbed on equivalent sites.
 The phase factor $ e^{i(\vec k-\vec k')\cdot \vec{ r}_l}$,  whose origin is discussed in  Appendix \ref{appendixA}, can be omitted in the case of a single Kondo impurity, where one can always take
 the origin of coordinates at the impurity site. In contrast, 
 for more than one impurity,  the phase factor   plays a central  role.  A hint of this comes from the following argument. 
  If we would ignore the phase factor, we could write: 
\beqa
{\cal V}&=&   J \vec{S}_T\cdot\vec s (0), 
\label{VKONDO2}
\eeqa
where $\vec{S}_T\equiv \sum_l\vec{S}(l)$ is the total spin of the chain, and $\vec s(0)$ is the spin density operator of the itinerant electrons at the origin. 
In the case of Heisenberg chains, where the single ion-anisotropy is neglected,  the total spin of the chain is a good quantum number. This implies that the Kondo coupling  in Eq. (\ref{VKONDO2})  could not  induce transitions between states that belong to manifolds with different total spin $S_T$.   As we discuss in detail below, this is far from being the case in experiments~\cite{Loth_Bergmann_natphys_2010}.  Therefore, the phase factors have to be included, 
a well known point~\cite{Jones_Varma_PRL_1988,Zachar96,Neto_Jones_PhysRevB_2000,Shah_Millis_PRL_2003,Lobos_Cazalilla_PRB_2012}  missed in recent papers dealing with Kondo interactions with short spin chains ~\cite{Rossier_prl_2009,Delgado_Rossier_prb_2010,Delgado_Rossier_ap_2012,Ternes_njp_2015,Gauyacq_Lorente_jp_2015}. .

\section{Bloch-Redfield approximation \label{BRF}}  
In order to treat the influence of the Kondo interaction on the dynamics of the 
spin chain, we adopt the standard Bloch-Redfield (BR) approach for open quantum systems weakly coupled to a reservoir~\cite{Cohen_Grynberg_book_1998}, which we briefly review here for completeness.   In order to implement the BR approach,  we first solve  exactly the atomic spin chain Hamiltonian 
\beqa
{\cal H}_{\rm chain}|M\rangle= E_M |M\rangle,
\eeqa
where $M$ labels the eigenstates $|M\rangle$.  The influence of the electron gas on the dynamics of the chain is treated  up to second order in the Kondo coupling ${\cal V}$.  The resulting approximate dynamical equation for the reduced density matrix $\hat\sigma(t)$ of the system is~\cite{Cohen_Grynberg_book_1998}: 
\beqa
\frac{d\hat\sigma(t)}{dt}= -\frac{i}{\hbar} \left[ {\cal H}_{\rm chain},\hat\sigma(t)\right]  + {\cal R}\left[\hat \sigma(t)\right].
\eeqa
The first term on the right hand side describes the coherent evolution of the states of the magnetic nanostructure, while the second is responsible of the dissipative dynamics.
  In general, the evolution of diagonal  terms in the density  matrix, $P_M\equiv \sigma_{M,M}$, referred as
  populations, and the off-diagonal terms,  known as coherences,  are coupled through the  BR equations.  However, there are many instances~\cite{Cohen_Grynberg_book_1998} where the dynamics of  populations
 and coherences are decoupled.   In that case, we actually write the so called Pauli master equation for the populations
\beqa
\dot{P}_{M}(t) =  -P_M(t)\sum_{M'} \Gamma_{M,M'} + \sum_{M'}P_{M'}(t)\Gamma_{M',M},
\eeqa
where $\Gamma_{M',M}$ are the population scattering rates. For a given transition, we define the spin relaxation time $T_1=\Gamma^{-1}_{M,M'}$.   In turn, 
the  dissipative dynamics of the coherence $\sigma_{MM'}$ of a pair of  non-degenerate states $M$ and $M'$ satisfies
\beqa
\dot{\sigma}_{M,M'}(t)
 = -i\frac{\Delta_{M,M'}}{\hbar} \sigma_{M,M'}(t) +  {\cal R}_{MM'MM'} \sigma_{M,M'}(t),
\nonumber
\eeqa
provided that no other couple of states $N$ and $N'$  has $\Delta_{N,N'}=\Delta_{M,M'}$, 
where $\Delta_{M,M'}=E_M-E_{M'}$. 
  If we write down
\beqa
{\cal R}_{MM'MM'} = -\gamma_{M,M'} -  i \delta\omega_{M,M'},
\label{2sides}
\eeqa
we can write the dynamical equation for the coherence as: 
\beq
\dot{\sigma}_{M,M'}(t)=
 -i\frac{\tilde{\Delta}_{M,M'}}{\hbar} \sigma_{M,M'}(t) + 
\gamma_{M,M'} \sigma_{M,M'}(t).
\label{cohren}
\eeq
Equation (\ref{cohren}) 
 describes both a decay of the coherence  on a time scale $T_2\equiv \gamma_{M,M'}^{-1}$ as well as a shift of the transition energy, 
 $\tilde{\Delta}_{M,M'}\equiv  \Delta_{M,M'}+\hbar\delta\omega_{M,M'}$.  Equations (\ref{2sides}-\ref{cohren}) clearly shows that decoherence and renormalization of the transition energy are intimately related, being the real and imaginary part of the same self-energy function~\cite{Delgado_Hirjibehedin_sci_2014} and thus, contain the same oscillatory dependence on $k_Fd$ displayed by the RKKY interaction~\footnote{To be more accurate, the shift corresponds to the dynamical or retarded version of the RKKY interaction}.
 Then, the dissipative dynamics has 3 types of consequences on the magnetic nanostructure:
\begin{enumerate}
\item    Scattering between states $|M\rangle$ to $|M'\rangle$, at a rate $\Gamma_{M,M'} \equiv T_1^{-1}$.
\item   Decay of the coherence on a time scale $ T_2$ between a pair of  eigenstates $|M\rangle,|M'\rangle$.
\item Renormalization of the energy levels.
\end{enumerate}

We emphasize that these three effects are totally general of quantum systems coupled to their environment.
For a single atomic spin~\cite{Delgado_Hirjibehedin_sci_2014}, they have been study theoretically.  Their magnitude is controlled by the dimensionless parameter $\rho J$, where $\rho$ is the density of states at the Fermi energy.  For the spin chains we shall demonstrate that, in addition,  $k_F d$ plays a central role.
In the case of the renormalization of the energy levels, this gives the very well known RKKY interaction~\cite{Kittel_Fong_book_1987}.  Interestingly,  these collective effects have been overlooked in the case of $T_1$ and $T_2$~\cite{Gauyacq_Lorente_jp_2015,Delgado_Rossier_prb_2010}.

A detailed derivation of the different rates appearing in the BR approach for a small arrays of spins can be found in the recent review~\cite{Delgado_Rossier_rev}. The general expressions of the BR tensor for the Kondo coupling (\ref{VKONDO}) are presented in Appendix \ref{appendixB} for completeness, while the particular expressions for $T_1^{-1}$, $T_2^{-1}$ and $\delta \omega_{M,M'}$ are given below.  

\subsection{Scattering rate $1/T_1$ \label{T1rates}}

Let us consider two eigenvectors $|M\rangle$ and $|M'\rangle$ of  ${\cal H}_{\rm chain}$.
The scattering rate from   $|M\rangle$ to  $|M'\rangle$  induced by the Kondo coupling  is given by 
\beqa
&&
\Gamma_{MM'}=
\frac{\pi J^2}{2\hbar^2} 
\sum_{\vec k,\vec k'}
f(\epsilon_k) \left(1-f(\epsilon_{k'})\right)
\crcr
&&\qquad \times
\chi_{M,M'}(\vec{k}-\vec{k}')
\delta\left(\epsilon_k+E_M-\epsilon_{k'}-E_{M'}\right),
\label{dinteg}
\eeqa
where
\beq
\chi_{M,M'}(\vec{q})\equiv 2\sum_{l,l'=1}^N
e^{i\vec q\cdot (\vec r_l-\vec r_{l'})}
\sum_aS^a_{MM'}(l)S^a_{M'M}(l') ,
\label{defchi}
\eeq
and 
$S^a_{MM'}(l)\equiv \langle M|S_a(l)|M'\rangle$ with $a=x,y,z$.
The physical interpretation of  Eq. (\ref{dinteg}) is quite transparent.  The factor $f(1-f)$ reflects that the scattering rate is proportional to the occupation of the initial quasiparticle state and the availability of the final quasiparticle state.
The delta function enforces the energy conservation of the whole process. 

Critically important, the $\chi$ function encodes several aspects that are essential in the rest of the paper.  It is given by the atomic spin matrix elements, summed over all the chain sites and weighted by the Bloch phase factors.  This function entails a spin sum rule: transitions are only permitted if the change in the atomic spin $S_z$  is either 0, or $\pm 1$,  which respects  the conservation of the total spin in the Kondo exchange interaction.  Formally, $\chi$ arises from the Fermi Golden Rule expression for the scattering rates, which contains the square of the perturbing Hamiltonian.  In our case, the perturbing Hamiltonian contains a sum over the atomic sites and the initial and final quantum numbers of the electron-hole pairs created in the Kondo scattering.

\subsection{Decoherence rate $1/T_2$  \label{T2chain}}
The Bloch-Redfield approach permits us to extract the decoherence rate $1/T_2$ between any two eigenstates of the isolated chain. The decoherence rate $\gamma_{M,M'}=1/T_{2}$  contains two contributions, the so called
adiabatic and the nonadiabatic terms~\cite{Cohen_Grynberg_book_1998}. The first comes from $T_1$-like  population scattering  processes~\cite{Cohen_Grynberg_book_1998}:
\beqa
\gamma^{nonad.}_{M,M'}=\frac{1}{2}\left(\sum_{N\ne M} \Gamma_{M,N}+\sum_{N\ne M'} \Gamma_{M',N}\right),
\label{gnonadv}
\eeqa
where $\Gamma_{M,M'}$ are the scattering rates defined in Eq. (\ref{dinteg}).

The adiabatic contribution or {\em pure dephasing}  corresponds to processes that occur even in the absence of changes in populations
of the $|M\rangle$ states. It is driven
by  elastic scattering processes with the reservoir.   In our case,  the adiabatic decoherence rate is given by:
\beqa
\gamma_{M,M'}^{ad.}&=&\frac{\pi J^2}{2\hbar}
\sum_{\vec k,\vec k'}
f(\epsilon_k) \left(1-f(\epsilon_{k'})\right)
\crcr
&&\qquad \times
\chi_{M,M'}^ {ad.}(\vec{k}-\vec{k}')
\delta\left(\epsilon_k-\epsilon_{k'}\right).
\label{invT2chain}
\eeqa
The matrix elements $\chi_{M,M'}^ {ad.}(\vec{q})$ are given by (see Appendix \ref{appendixB} for details):
\beq
\chi_{M,M'}^ {ad.}(\vec{q})=\sum_{a} \left|\sum_{n=1}^N \left(e^{i\vec q\cdot \vec r_n}S^a_{MM}(n)-e^{-i\vec q\cdot \vec r_n}S^a_{M'M'}(n)\right)\right|^2.
\label{chiadb}
\eeq

\subsection{Renormalization  of the energy levels \label{shifts}}
The imaginary part of the BR tensor
 gives place to an effective system Hamiltonian that commutes with ${\cal H}_{\rm chain}$~\cite{Breuer_Petruccione_book_2002} and hence, it is diagonal in the $\left\{|M\rangle\right\}$  bases. Thus, the only possible non-zero contributions to the energy shifts are given by the components of the form ${\cal R}_{NMNM}$. The variation of the bare frequencies $\delta \omega_{MM'}$ can then be decomposed into the shifts of single levels, i.e., $\hbar \delta\omega_{MM'}=\delta E_M -\delta E_{M'}$. For the Kondo interaction (\ref{VKONDO}), this energy shifts takes the form (see Appendix \ref{appendixB} for the details):
 \beq
 \delta E _{M}=\frac{J^2}{\hbar}{\cal P} \sum_R\sum_{\vec k\vec k'} 
\frac{f(\epsilon_{\vec k})\left(1-f(\epsilon_{\vec k'})\right)}{\omega_{\vec k\vec k'}+\omega_{MR}}
\tilde \chi_{MR}(\vec k-\vec k'),
\label{shiftRKKY}
 \eeq
where ${\cal P}$ stands for the principal part and the matrix elements $\tilde\chi_{MR}(\vec q)$ are given by 
\beqa
\tilde \chi_{MR}(\vec q)=2\sum_{a} 
\left| \sum_{n=1}^N  S^a_{MR}(n)e^{i\vec q \cdot \vec r_n}
\right|^2.
\eeqa
Application of this equation for a  single anisotropic spin  results in the renormalization of the single spin anisotropy due to Kondo exchange~\cite{Oberg_Calvo_natnano_2013,Delgado_Hirjibehedin_sci_2014} observed experimentally~\cite{Oberg_Calvo_natnano_2013,Jacobson_Herden_natcom_2015}.
Equation (\ref{shiftRKKY}) can be formally connected with the  conventional RKKY formulas if we replace the matrix elements $ S^a_{MR}(n)$ the $a$ component of a classical magnetic moment at atom $n$, and we take the static limit where $\omega_{MR}=0$.

\section{Spin relaxation in spin dimers\label{secdimer}}
We now apply the BR formalism to compute $T_1$ for the simplest spin array: an spin dimer.
The results can be readily extended to longer chains, but the essential new physics
appears already at the 2-spin level. For reference,  we start by revisiting the single spin case~\cite{Delgado_Rossier_prb_2010,Delgado_Hirjibehedin_sci_2014}.

\subsection{The single spin case\label{T1mono}}
By introducing the density of states
$\rho(\epsilon)=\sum_{\vec k,\sigma}\delta\left(\epsilon_{\vec k\sigma}-\epsilon\right)$, we can write Eq. (\ref{dinteg}) for a single spin located at $\vec r=0$ as
\beqa
&&
\Gamma_{MM'}=
\frac{\pi J^2}{4\hbar}
\iint \!\rho (\epsilon) \rho(\epsilon') f(\epsilon) \left(1-f(\epsilon')\right)
\crcr
&&\times
\sum_a |S_a^{MM'}|^2 
\delta\left(\epsilon-\epsilon'+\Delta_{MM'}\right) d\epsilon d\epsilon'.
\label{dinteg1}
\eeqa
For a single spin, the matrix elements  $|S_a^{MM'}|^2 $ only connect states with components of $S_z$ that differ, at most, in one unit, which translates into the selection rule $\Delta S_z=0,\pm 1$ observed in inelastic electron tunneling spectroscopy (IETS)~\cite{Rossier_prl_2009}.  
We now make the additional assumption $\rho(\epsilon)\approx \rho$ in the energy interval giving the dominant contribution to (\ref{dinteg1}), which is of the order of $k_BT$ around the Fermi level.
 We thus obtain the following result~\cite{Delgado_Rossier_prb_2010}:
\beqa
\Gamma_{MM'}=
\frac{\pi (\rho J)^2}{4\hbar}  {\cal G}(\Delta_{MM'})
\sum_a |S^a _{MM'}|^2 
\label{dinteg1b}
\eeqa
where
${\cal G}(\Delta)\equiv \Delta\left(1+ n_B(\Delta)\right)$,
with $n_B(x)$ the Bose occupation factor. In the case of relaxation of an excited state $M$, where $\Delta_{MM'}>0$, there are two interesting limits. First,  $\Delta\gg k_BT$ in which case ${\cal G}(\Delta)\simeq \Delta$, and the relaxation rate is proportional to the energy difference.  In the opposite limit where the splitting is much smaller than $k_BT$,  we get that ${\cal G}\simeq k_BT$.

Let us now consider the spin relaxation of the lowest energy excitation of a spin $S$ with easy axis anisotropy $DS_z^2$ ($D<0$) at zero external field. In such a case, the ground state of the system has $S_z=\pm S$ and the first excitation corresponds to $S_z=\pm(S-1)$. The decay rate is then given by
\beq
\Gamma_{(S-1)\to S}=
\frac{\pi (\rho J)^2}{4\hbar}  {\cal G}(\Delta) S,
\eeq
where $\Delta=(E_{S-1}-E_{S})= (2S-1)|D|$.
 For instance, if we consider the $S=5/2$ case, relevant for the experimental case of a Mn adatom on a Cu$_2$N/Cu(100) surface~\cite{Hirjibehedin_Lin_Science_2007}, 
  direct scattering between the two degenerate ground states would {\em not} be possible.  In contrast, if we add the in-plane anisotropy term $H= -|D| S_z^2+ E(S_x^2-S_y^2)$, direct scattering is permitted~\cite{Delgado_Rossier_prl_2012}.  

For convenience, we introduce the relaxation rate $\Gamma_0(\Delta)$ of a $S=1/2$ spin with a 
Zeeman splitting $\Delta$ 
\beq
\Gamma_0(\Delta)=
\frac{\pi (\rho J)^2}{8\hbar}  {\cal G}(\Delta). 
\label{GammaZ}
\eeq
Unless otherwise stated, we refer all the rates  for spin dimers to the monomer rate  $\Gamma_0(\Delta)$. For reference, if we take  $\rho J=0.1$   we have $\Gamma_0(\Delta=1 meV)\simeq 6\, GHz$.

\subsection{A Heisenberg spin dimer \label{dimerT1}}

We now consider the simplest possible chain,  a dimer described by the following Hamiltonian: 
\beq
{\cal H}_{\rm dimer}= |J_H|\vec{S}(1)\cdot\vec{S}(2) 
+g \mu_B  B_z \left(S_z(1)+ S_z(2)\right).
\label{Hdimer}
\eeq
The model can be solved analytically taking into account that $[{\cal H}_{\rm dimer}, S_T^2]=[{\cal H}_{\rm dimer}, S_Z]=0$, where $S_T$ and $S_Z$ are the total spin operator and its third component.
Thus, the eigenvectors of $S_T^2$ and $S_Z$, $|S_T,S_Z\rangle$, are also eigenvectors of ${\cal H}_{\rm dimer}$.
For simplicity, hereafter we discuss a homogeneous dimer with  $S(1)=S(2)\equiv S$. Although Hamiltonian (\ref{Hdimer}) assumes an antiferromagnetic coupling, the main conclusions of this section equally apply to the ferromagnetic case.

We consider two limits of this model. First, the $B_z=0$ limit,  where
the ground state is the spin singlet $|0,0\rangle$ and the first excited state is the spin triplet, with excitation energy $|J_H|$, see Fig. \ref{fig2}(a). We focus on the relaxation of this first excited state to the ground state.   
The second limit of interest is for very large $B_z$ and  two $S=1/2$ spins, so that the ground state $|1,-1\rangle$ is unique, while there are two possible excited states that can decay directly to the ground state, the $|1,0\rangle$ triplet and the $|0,0\rangle$ singlet.
  We shall see that these two transitions are the spin analogue of the  sub-radiant and super-radiant case  for electromagnetic emission of a couple of two-level atoms~\cite{Cohen_Grynberg_book_1998}.

\subsection{Relaxation rate of a spin dimer}
\subsubsection{Evaluation of $\chi_{M,M'}(\vec{q})$}
In order to evaluate $\chi_{M,M'}(\vec{q})$ for the dimer, 
we write $\vec{r}_1-\vec{r}_2= d\hat{n}$, where $d$ is the interspin distance distance and $\hat{n}$ is the unitary vector along the line that joins the two atoms of the chain.  This permits us to write down the $\chi$ function of the dimer as: 
\beqa
\chi_{M,M'}(\vec{q})&\equiv & 2 \sum_{a,l} |S_a^{M,M'}(l)|^2 
\crcr
&+&4{\rm Re}\Big[e^{i d \vec{q}\cdot \hat{n}}\sum_a S_a^{MM'}(1)S_a^{M'M}(2)  \Big].
\eeqa 
The  first line  represents the 
scattering rate of the dimer as a sum of two independent monomers 
(although
the spin wave functions are correlated). Thus, it corresponds to the incoherent sum of the contributions coming from each atomic spin.  The second line accounts for the interference, where scattering with two atoms at the same time occurs.

An special value of this function is
$\chi_{M,M'}(0 )\equiv 2 \sum_{a} | \langle M|(\hat S_a(1)+\hat S_a(2))|M'\rangle |^2$.
Hence, since the Heisenberg model commutes with the total spin,  for any two eigenvectors of $S_Z$, $|M\rangle$ and $|M'\rangle$ that  correspond  to different eigenvalues of $S_T^2$, we have $\chi_{M,M'}(0 )=0$.

At this point, the crucial role played by the Bloch phases is apparent. The sum over momentum in Eq. (\ref{dinteg}) translates essentially into an integral in a tiny energy window around  the Fermi surface. Therefore, the phase factors will depend strongly on the geometry of the Fermi surface.    In the following we show this in the  case of a single parabolic band. We consider  independent electrons in
one, two and three dimensions, for which    the Fermi surface is made of two points (1D), a circle (2D) or a sphere (3D).  Since our initial motivation is the study of adatoms, the  case of 2D and 3D electrons is closer to describe experiments. However, the 1D leads to analytical expressions and provides insight on the phenomenon of multi-site scattering addressed here.

\begin{figure}[hbt]
\includegraphics[width=1.\linewidth]{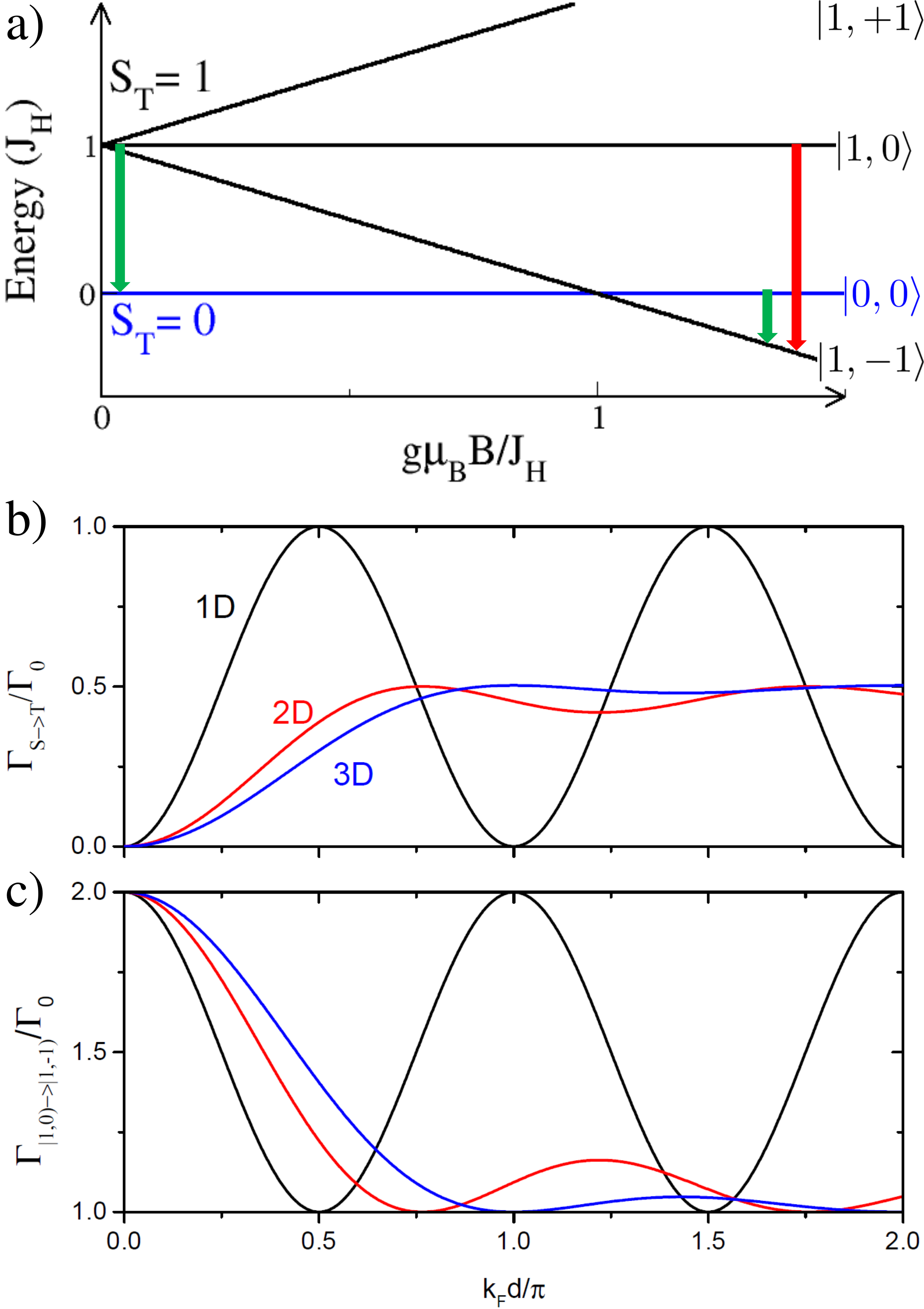}
\caption{ \label{fig2}(Color online).
(a) Scheme of the energy levels of a $S=1/2$ Heisenberg spin dimer versus the applied magnetic field $B$.
The vertical arrows mark the two transition discussed in the main text, the singlet-triplet transition (green arrows), displayed in panel (b), and the triplet-singlet (red arrow), displayed in panel (c).
 (b) Variation of the decay rate between the excited triplet state and the singlet ground state at zero field with $k_Fd$.
(c) Decay rate between the $|1,0\rangle$ triplet excited state and the $|1,-1\rangle$ ground states at large magnetic field, $g\mu_BB_z=10k_BT$. In all cases, $J_H=5k_BT$. Notice the different scales on the vertical axes.
}
\end{figure}

\subsubsection{$T_1$ for the dimer:  coupling to 1D fermions\label{sub1d}}
For 1D fermions it is always possible~\cite{Leggett_Chakravarty_rmphys_1987} to  linearize the bands around the two Fermi points, $\pm k_F$, 
$\epsilon(k) = \pm \hbar v_F (k\mp  k_F)\equiv \pm \hbar q$.  Given that only the contributions at the neighborhood of the Fermi energy are relevant, for any smooth function $g(k)$, we can replace 
\beqa
&&\hspace{-1.cm}\int_{-\infty}^{\infty} f(\epsilon_k)\left(1-f(\epsilon_k+\omega)\right)g(k) \frac{dk}{2\pi} 
\crcr
&& \approx\quad\rho \sum_{s=\pm 1} g\left(s k_F\right)\int_{-\infty}^\infty  f(\epsilon) \left(1-f(\epsilon+\omega)\right)d\epsilon,
\eeqa
where $\rho=(2\pi\hbar v_F)^{-1}$.
The original double integral becomes thus the sum of 4 terms, two of which have $k-k'$ very small, and the other two with $k-k'\simeq \pm 2 k_F$. In both cases we can take $\chi(k-k')$ out of the integral, so that we obtain: 
\beqa
&&\hspace{-0.5cm}\Gamma_{MM'}=
\frac{\pi \left(\rho J\right)^2}{32\hbar}{\cal G}( \Delta)
\crcr
&&\hspace{-0.1cm}\times
2 \left( \chi_{M,M'} (0) + \chi_{M,M'}( 2k_F d) \right).
\label{dinteg1Db}
\eeqa
 Thus, the relaxation rate due to interaction with 1D fermions can be written as the spin relaxation of an individual spin multiplied by a structure factor that has contributions coming from forward and backward scattering,  given by $\chi_{M,M'}(0)$ and $\chi_{M,M'}(2 k_F d)$ respectively.

We now work out these general formulas for the  case of two $S=1/2$ spins at zero field.  Since we consider an antiferromagnetic dimer, the excited state is actually a triplet with $S_T=1$, and the  ground state is the singlet with $S_T=0$. After some  simple algebra 
 we obtain:
\beqa
\chi_{|1,S_Z\rangle\to |0,0\rangle}(q)=\left(1- \cos(qd)\right)=2\sin^ 2\left(\frac{qd}{2}\right).
\label{25simpl}
\eeqa 
This results anticipates an oscillatory dependence of the spin relaxation of the dimer on the dimensionless parameter $k_F d$, which evokes the oscillations in the RKKY coupling.
Moreover, we have $\chi_{|1,S_Z\rangle\to |0,0\rangle}(0)=0$. 
 This is expected since, for  $q=0$, the Kondo Hamiltonian commutes with the atomic spin operator $\hat S^2_T$. As a result, the representation of the Kondo operator in the basis of eigenstates of $\hat S_T^2$ is diagonal, i.e., there are no transitions between states with different $S_T$.   This highlights the crucial role played  by the Bloch phases in the Kondo Hamiltonian, regardless the dimensionality.  In the specific case of a bath of 1D fermions, the fact that $\chi_{|1,S_Z\rangle\to |0,0\rangle}(0)=0$ 
 implies that only backward scattering events contribute to the atomic spin relaxation of the dimer, see Eq. (\ref{dinteg1Db})  
  
After the substitution of Eq. (\ref{25simpl}) into (\ref{dinteg1Db}), we get that the 
 relaxation rate between any of the triplet excited states and the singlet ground state is given by
\beqa
\Gamma_{|1,S_Z\rangle \to |0,0\rangle}= 
 \Gamma_0(\Delta_{ST})\left(1-\cos^ 2\left(k_F d\right)\right)
\label{gammast}
\eeqa
where $\Delta_{ST}=J_H$ and  $\Gamma_0(\Delta)$ is given by Eq. (\ref{GammaZ}). 

Equation (\ref{gammast}) is the simplest example of the main result of this work: the spin relaxation of a coupled-spin systems driven by Kondo exchange has an RKKY-like oscillating dependence on $k_Fd$, as well as the scaling of the rate proportional to $(\rho J)^2$.  
As anticipated, the coupling of the quantum spin chain to a reservoir (the conduction electrons) results both in a finite lifetime of the excited states ($T_1$) and a renormalization of the energy levels, both oscillating functions of $k_Fd$.

 \subsubsection{Subradiant and superradiant states}
 We now consider the problem of the $S=1/2$ dimer in the limit of   large field, i.e., $B_z\gg J_H/g\mu_B$. In this limit, the ground state corresponds to the state $|1,-1\rangle=\left|\downarrow\downarrow\rangle\right.$. 
The two  excited states that can relax into the ground state via Kondo interaction are  the singlet $|0,0\rangle$ and triplet state $|1,0\rangle$, 
  with excitation energy $g\mu_B B_z -J_H$ and $g\mu_B B_z$ respectively, see Fig.~\ref{fig2}a).
    Their wave functions are 
 \begin{eqnarray}
 |1,0\rangle&=& \frac{1}{\sqrt 2}\left(|\uparrow\downarrow\rangle +  |\downarrow\uparrow\rangle\right) \nonumber\\
  |0,0 \rangle&=& \frac{1}{\sqrt 2}\left(|\uparrow\downarrow\rangle -  |\downarrow\uparrow\rangle\right). 
  \end{eqnarray}
   The relaxation takes place via emission of an $S_Z=1$ electron-hole pair in the Fermi liquid. The relaxation of state  $\left|\uparrow\uparrow\rangle\right.$ requires a change in $S_Z$ of two units, and therefore can not decay to the ground state through emission of a single electron-hole pair.  
   
 At this point, it is convenient to draw the analogy between this system and a couple of 2-level atoms coupled to the photon vacuum, where each Zeeman-split spin behaves as a two-level atom.   In the case of the 2-level atoms, the relaxation would take place through a photon emission with $\Delta S_z=\pm 1$.  
Importantly,  the  two different excited states, $|1,1\rangle$ and $|0,0\rangle$,  are mathematically identical 
 to the so called subradiant and superradiant states in quantum optics~\cite{Dicke_pr_1954,Cohen_Grynberg_book_1998}.  As a result, 
 we expect that in the limit of $k_F=0$,  analogous to the limit in which the  wavelength $\lambda$ is much larger than the interatomic separation $d$,  the relaxation rates of these two states should be radically different.  
 
   The transition between the excited singlet state and the triplet ground state is given essentially by Eq. (\ref{gammast}) where now $\Delta'=g\mu_B B_z-J_H>0$. 
   By contrast, the decay rate from the excited triplet state to the ground state is given by
\beqa
\Gamma_{|1,-1\rangle\to |1,0\rangle}=\Gamma_0(g\mu_B B_z)\left(1+\cos^2(k_Fd)\right).
\label{gammatt}
\eeqa
For  $k_Fd=0$,  we have that the large field-singlet excited state does not decay (the expression would be identical to that in Eq. (\ref{gammast}),  but with the singlet triplet splitting modified, see Fig.~\ref{fig2}(a) ).
 However, the triplet $|1,0\rangle$ decays twice as fast as two independent single spins would do, in exact analogy with the concept of subradiant (singlet) and super-radiant (triplet) states proposed originally in the context of quantum optics, and adapted to the case of spin relaxation coupled either to photons~\cite{Chudnovsky_Garanin_2002} or phonons~\cite{Chudnovsky_Garanin_2004}.

 The physical origin  of this phenomenon is the interference between the two channels for the decay. In the photon language,  the emission can take place from one of the two atoms and, when the wavelength of the photon $\lambda$ is much larger than the interatomic distance, these two channels are indistinguishable.   In the singlet case, the excited state wave functions of the two atoms are completely out of phase. So, the net dipole moment vanishes, and there is no emission. In the spin language, the total spin vanishes, and hence the interaction with electrons bath vanishes.  For the triplet, the wave functions of the two atoms are in phase, building a larger electric dipole that results in a faster radiative decay.  This discussion highlights the role played by the spin correlations driven by intra-chain interactions on the spin relaxation rate.   

We now discuss the effects associated to the finite value of $k_F$, that would correspond in the analogy with quantum optics to the case where $\lambda$ is no longer much larger than the interatomic distance.   
In the case of  1D fermions, the extremely simple geometry of the Fermi surface gives a quite simple result: since  relaxation is only induced by backward scattering events, 
 the contributions from each spin in the dimer are dephased by $2k_F d$.
 The peculiarities of the present result associated to the one dimensional character of the Fermi gas will be highlighted in the following section.

\subsubsection{$T_1$ for the dimer:  coupling to $2D$ and $3D$  fermions \label{3dcase}}

So far, we have  considered the relaxation induced by scattering with  1D fermions for which the Fermi surface is made of two points, leading
to spin relaxation rates that present undamped oscillations as a function of  $k_Fd$.  
We now consider the case of 2D and 3D fermions in a  parabolic band.  The main difference with the 1D case is the fact that there is now a continuum of possible 
quasi-elastic scattering events.
 Using the general expressions (\ref{dinteg}) and (\ref{defchi}),   the integration over the angular degrees of freedom of the electron bath can be done analytically. After some simple algebra, we obtain  the following expressions for the decay rate between the triplet and singlet states at zero field:
\beqa
\Gamma_{|1,S_Z\rangle \to |0,0\rangle}^d
= \Gamma_0(\Delta_{S,T}){\cal F}^d(k_Fd)
\label{gammastd}
\eeqa
where
\beqa
 {\cal F}^d(x)=\left\{
 \begin{array}{ll}
 \left(1-J_0^2(x)\right), & d=2\\
 \left(1-{\rm sinc}^2(x)\right), &d=3
 \end{array}
 \right.
\eeqa
with $J_0(x)$ the zero order Bessel function and ${\rm sinc}(x)=\sin x/x$. Similarly, the decay rate $\Gamma_{|1,0\rangle\to |1,-1\rangle}$ will be given by
\beqa
\Gamma_{|1,-1\rangle\to |1,0\rangle}^ d=\Gamma_0(g\mu_B B_z){\cal F}_{T}^d(k_Fd)
\label{gammattd}
\eeqa
 with 
\beqa
 {\cal F}_T^d(x)=\left\{
 \begin{array}{ll}
 \left(1+J_0^2(x)\right), & d=2\\
 \left(1+{\rm sinc}^2(x)\right), &d=3
 \end{array}
 \right. .
\eeqa

The variations of the transition rates of Eqs. (\ref{gammastd}) and (\ref{gammattd}) with $k_Fd$  are plotted in Fig. \ref{fig2}b)  and c) respectively. 
We find  that, in 2D and 3D  the transition rates are also oscillating functions of $k_Fd$, but the amplitude of the oscillation decays as $k_Fd$ increases, making the result even more similar to the RKKY interaction.  For $k_Fd=0$ we still have the perfect cancellation of the relaxation rate in the subradiant state and the enhancement, by a factor of 2, of the superradiant state. Interestingly,  in the limit where the oscillations are damped, the sub-radiant state still emits at a rate half the one of an individual spin, whereas the superradiant state emits twice as much as the subradiant one.    The origin of the damped oscillations can be traced back to the interference between many quasi-elastic scattering events at the Fermi surface, picking  different phases when scattering with different atoms.

\subsection{Spin dimer: comparison with experiments\label{expdiss}}
The previous discussion solves a formerly identified puzzle~\cite{Delgado_Rossier_prb_2010,Delgado_Rossier_ap_2012} in the interpretation of existing experiments of antiferromagnetically coupled Mn dimers on Cu$_2$N~\cite{Hirjibehedin_Lutz_Science_2006,Loth_Bergmann_natphys_2010}.
Although we have centered our discussion of the spin dimer on the $S=1/2$ case, exactly the same treatment can be done for the $S=5/2$ case, relevant for the Mn dimer. In particular, the expression for the spin relaxation between the $S=1$ excited state towards the  $S=0$ ground state
is identical to Eq. (\ref{gammast}).   Ignoring the Bloch phases in the Hamiltonian, as it has been done in all previous works in this field~\cite{Rossier_prl_2009,Delgado_Rossier_prb_2010,Gauyacq_Lorente_pss_2012,
Ternes_njp_2015},  
leads to the conclusion that the spin relaxation for the dimer due to Kondo interactions is forbidden.  
However, the current-dependence of the non-equilibrium  $dI/dV$ profiles of the Mn-dimer permits one to extract relaxation times of the excited states in the range of $3$ and $50$ ps for the first and second excitations respectively~\cite{Loth_Bergmann_natphys_2010}.

 An even more compelling evidence comes from the lack of modulation of the inelastic excitation signal as the STM tip is moved laterally along the dimer. In the ideal situation in which the tip is coupled to only one atom, 
 there is no possible destructive interference and a clear step is expected.  When the tip is right in the middle of the two atoms, a theory ignoring the phase difference between surface single particle states scattering with the two atoms would predict a vanishing transition rate~\cite{Delgado_Rossier_ap_2012}. In contrast,  such a modulation has not been observed experimentally.  
   Of course, once the Bloch phases are restored in the Hamiltonian,  this cancellation is no longer there. Actually, 
if we consider the bulk Fermi energy of Cu, $E_F\approx 7$ eV~\cite{Ashcroft_Mermin_book_1976} and $k_F\approx 1.36 $ \AA$^{-1}$. Since, from the STM topography, the estimated distance between the Mn is $d\approx 3.6$ \AA, we get that $k_Fd\approx  1.56\pi$ and $\sin^2(k_Fd)\approx 1$, in perfect agreement with the fully developed inelastic step in the IETS~\cite{Loth_Bergmann_natphys_2010}.

\section{Spin relaxation rate of spin-wave states in short ferromagnetic chains\label{SW_ferro}}

Motivated by recent experiments~\cite{Spinelli_Bryant_naturematerials_2014}  in  short chains of Fe atoms on 
the Cu$_2$N/Cu(100)  surface, which behaves as anisotropic chains of ferromagnetically coupled $S=2$ spins,
we now consider the relaxation of spin wave excitations in short ferromagnetic (FM) chains.  The experiments~\cite{Spinelli_Bryant_naturematerials_2014} demonstrated that the amplitude of the spin excitations driven by tunneling electrons was modulated for different atoms in the chain. 
Moreover, the current-driven switching dynamics between the two ground states is also controlled by these rates~\cite{Spinelli_Bryant_naturematerials_2014}. 
Importantly, the excellent agreement between the theory, based on the exact solution of the anisotropic Heisenberg Hamiltonian  (\ref{hamilC}), and the experimental results, further validates the use of model Hamiltonians to describe nanoengineered spin chains. 
 The observed modulation could be associated to variations in the matrix element $|\langle G|\sum_{a} S_a(i)|SW\rangle|^2$ for different atoms $i$ in the chain, where $|G\rangle$ stands for one of the ferromagnetically align ground state and $|SW\rangle$ for the spin-wave excited states.

In this section we undertake a systematic study of the lifetimes of spin waves depending on
 3 factors: chain length $N$, the phase factor $k_Fd$, 
  and the modulation of the spin waves across the chain.

\subsection{Description of the spin wave states}
Here we briefly describe the spin wave states whose lifetimes we are interested in. For the sake of simplicity, we consider first the case of $S=3/2$ chains with $E=0$, 
so that the Hamiltonian commutes with the total  $S_z$. For the numerical evaluations we take $J_H=D=-1$ meV.
The doubly-degenerate ground state has the form
\begin{equation}
|G_{\pm}\rangle= |\pm\frac{3}{2}\rangle\otimes|\pm\frac{3}{2}\rangle ...... |\pm\frac{3}{2}\rangle.
\end{equation}
 Without loss of generality, from the two degenerate ground states at zero field, with $S_z=\pm 3N/2$, we choose the one with $S_z>0$ and focus on its spin wave excitations.  In the absence of applied magnetic field,  every state that we discuss here has a time-reversal degenerate partner. 
 
\begin{figure}
\includegraphics[width=1.\linewidth]{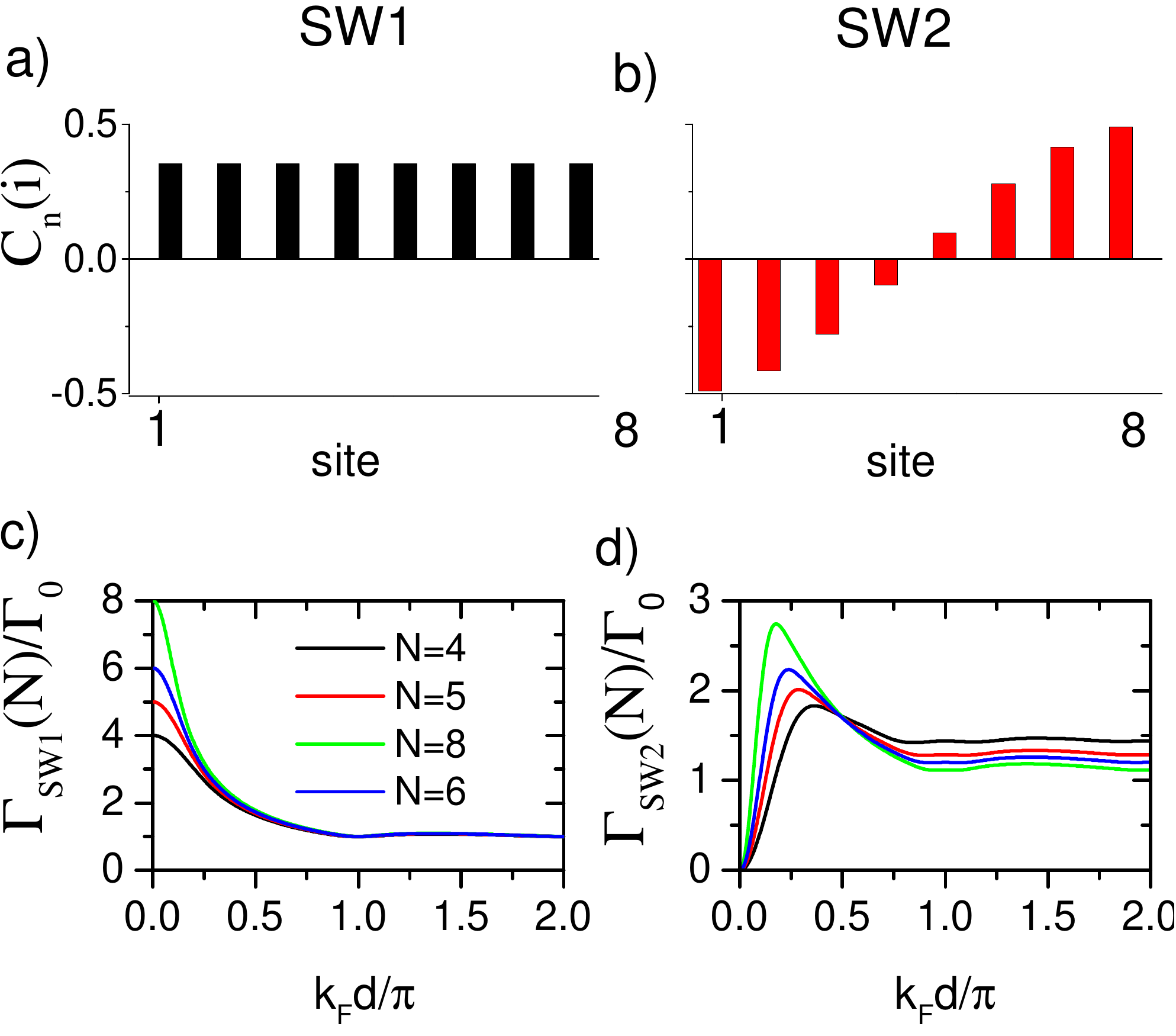}
\caption{ \label{fig3}(Color online). Lowest spin waves of an $S=3/2$ FM Heisenberg chain. 
 Coefficients $C_n(i)$ of the SW1 (a) and SW2 (b) for an $N=8$ chain. Relaxation rates of the SW1 [panel (c)]  and SW2 [panel (d)]. In all cases, a coupling to a three dimensional electron gas was assumed and $J_H=D=-1$ meV.  
}
\end{figure}

  Our simulations show that, regardless of the size of the chain, the first two excitations are spin waves (SW) with total spin $S_z=3N/2-1$ whose wavefunctions are linear combinations of the form 
\begin{equation}
|{\rm SWn}\rangle = \sum_{i=1,N} C_n(i) S^{-}(i)|G_+\rangle ,
\end{equation}
where $S^{-}(i)|G_+\rangle$ corresponds to an state in which the $i$-spin has flipped from $3/2$ to $1/2$, and
$C_n(i)$ are coefficients that are obtained from the exact diagonalization of the chain Hamiltonian. 
Interestingly, for arbitrary spin $S$ and chain length $N$, the first SW, denoted as SW1, always appears at $(2S-1)|J_H|$ above the ground state. This SW has an equal probability of flipping any of the spins in the chain, see Fig.~\ref{fig3}a).

Very close in energy, at $(2S-1)\xi|J|$, with $\xi\equiv \xi(N,S)\approx 1$ for $N\gtrsim 6$,  one finds that in addition to the SW1, there is a second excitation corresponding to another spin wave of different nature, which we denote by SW2. As SW1, it corresponds to a linear combination of states where one local spin $S_z(l)=3/2$ is switched to $S_z(l)=1/2$, but now with growing weights towards the border of the chain, see Fig.~\ref{fig3}a).  

\subsection{Spin-wave relaxation rates}
We now discuss the dependence of the spin relaxation of $|{\rm SW1}\rangle$ and $|{\rm SW2}\rangle$ both, on the size of the chain $N$ and on $k_F d$.  The transition rate for the SW1 mode towards the ground state, $\Gamma_{\rm SW1}(N)$, is shown in Fig.~\ref{fig3}b). Interestingly, at $k_Fd=0$ the following relation applies
 \begin{equation}
 \Gamma_{\rm SW1}(N)/\Gamma_{\rm SW1}(1) = N.
 \end{equation}
 In other words, the relaxation rate scales linearly with the number of atoms. This is again analogue to the super-radiance phenomenon. The incoherent emission of an electron-hole pair is enhanced when the spin relaxation due to scattering with conduction electrons at different sites of the chain occurs in phase. 
 As the electrons are able to resolve the individual spins, this collective enhancement of the relaxation starts to fade away. In fact, in the cases of coupling to a two and three dimensional electron gases, the spin-wave relaxation occurs  at the same rate that the relaxation of a single spin excitation with the same energy, $\Gamma_{\rm SW1}(N,k_Fd)\to \Gamma_{\rm SW1}(1)$ when $k_Fd\gg \pi$.

The situation is radically different in the case of SW2, see Fig.~\ref{fig3}c). When the Fermi wavelength is not capable of resolving the spin-chain structure, and thus the scattering of electrons with the local spins occurs in phase, SW2 can not decay to the ground state. However, this protection of the SW2 disappears quite fast for $k_Fd>0$, with a maximum relaxation occurring at $k_Fd\approx 0.2-0.4\pi$ for $4\le N\le 10$. By contrast, for very small Fermi wavelengths ($k_Fd\gg \pi$), the relaxation is equivalent to the single spin relaxation with the same excitation energy (notice that for SW2, the excitation energy is size-dependent).

\section{Calculation of $T_2$ for spin chains\label{T2_deco}}

We now discuss the pure spin decoherence time of adatoms chains due to Kondo interactions with the substrate.   
For reference, we revisit the single spin case~\cite{Delgado_Rossier_prl_2012}.

\subsection{Spin decoherence of a single spin}
We start by considering the decoherence rate of a single Kondo impurity. Applying the general expressions (\ref{invT2chain}-\ref{chiadb}), and after introducing the (constant) density of states, one gets that the adiabatic decoherence rate $\gamma_{MM'}^{\rm ad.}$ between two eigenstates $M$ and $M'$ of the single spin Hamiltonian is given by
\beqa
\gamma^{\rm ad.}_{M,M'}=\frac{\pi (\rho J)^2}{8\hbar}k_BT\chi^{{\rm ad.}}_{MM'},
\label{chiadb1}
\eeqa
where
$\chi^{{\rm ad.}}_{MM'}=\sum_{a}\left|S^a_{MM}-S^a_{M'M'}\right|^2$.
The matrix elements $\chi^ {\rm ad.}_{MM'}$ takes a particularly simple form in some cases. For instance, in the case of 
half-integer easy-axis spins described by Hamiltonian (\ref{hani}), if $M$ and $M'$ are the doubly degenerate ground state that for $E=0$ are eigenstates of $S^2,S_z$, we get that $\chi_{+S,-S}^ {\rm ad.}\approx 4S^ 2$~\cite{Delgado_Rossier_prl_2012}, which leads to the adiabatic decoherence rate 
\beqa
\gamma^{\rm ad.}_{M,M'}=\frac{\pi (\rho J)^2}{2\hbar}k_BT
S^2
\label{gadiabatic}
\eeqa
that scales with   $S^2$. 
For instance, taking sensible values of $\rho J=0.1$, $S=1$, and $k_BT=100 $ mK,  the decoherence time would be $T_2\sim 5 $ ns.

\subsection{The Ising limit for anisotropic  spin chains}
The Ising spin chain corresponds to an experimentally relevant limit for certain magnetic adatoms for which 
there is a  strong easy axis anisotropy ($-D \gg |J_H|$). As a result,  it is  possible to truncate Hilbert space of dimension $2S+1$  of each atomic spin $S$  and retain only the lowest energy doublet, whose wave functions are mostly made of $S_z=\pm S$~\cite{Delgado_Loth_epl_2015,Jia_Banchi_arXiv_2015,banchi2016}.  Within this subspace, both the Kondo and Heisenberg  the spin-flip interactions are suppressed,  resulting in an effective Ising coupling. 
Hereafter we will denote as $\hat\tau_{a}$  the Pauli matrices acting in this $2\times 2$ subspace.

In the case of integer spins described by Hamiltonian (\ref{hani}), there is a finite splitting between the ground state $|g\rangle$ and the first excited state $|x\rangle$, that we  denote by $\Delta_1$.  We thus can write the effective single spin Hamiltonian as 
 ${\cal H}_0 \simeq \Delta_1 \hat\tau_z/2$. 
Moreover, when the spin operator is represented in the subspace of the lowest energy doublet one gets~\cite{Delgado_Loth_epl_2015,Jia_Banchi_arXiv_2015}  
  \begin{equation}
 \vec{S}= (0,0,\langle g|S_z|x\rangle\hat \tau_x). 
 \end{equation} 
 
Thus, the resulting spin chain Hamiltonian  is nothing but the widely studied quantum Ising model in a transverse field~\cite{Delgado_Loth_epl_2015}:
 \beqa
 {\cal H}^{{\rm Ising}}&=&\frac{\Delta_1}{2} \sum_{l}\hat\tau_z(l)
+\frac{j_I}{4}\sum_{l}\hat\tau_x(l) \hat\tau_x(l+1),
 \label{HIsing0}
 \eeqa
where $j_I= 4J_H|\langle g|S_z|x\rangle|^2$. 
Interestingly, the  transverse field term ($\propto \Delta_1$) comes from the quantum spin tunneling of the individual magnetic atoms~\cite{Delgado_Loth_epl_2015}, although it can be modulated with an applied field~\cite{banchi2016}.
This Ising Hamiltonian  describes the transition between the quantum behavior, that dominates for small $N$,
and a classical one with two degenerate ground states~ \cite{Delgado_Loth_epl_2015}. In the quantum regime, it presents a non-degenerate ground state linear combination of states with finite atomic spin magnetization, so that the average local magnetization is zero. This clearly contrast with the classical behavior characterized by a finite local spin magnetization along the Ising coupling.  This phenomenology is compatible with the experimental observations of Loth {\em et al.} for  Fe chains on Cu$_2$N/Cu(100)~\cite{Loth_Baumann_science_2012}.
 
In the case of half integer spins at zero applied field, Kramers' theorem ensures, at least, a two-fold degeneracy of the ground state for the single spin Hamiltonian. We denote the states in the doublet as $|g_1\rangle$ and $|g_2\rangle$ 
and we choose these states to diagonalize the $S_z$ operator.
Within that basis, the effective spin chain Hamiltonian reads as
 \beqa
 {\cal H}^{{\rm Ising}}=
 \frac{j_I'}{4}\sum_{l}\hat\tau_z(l) \hat\tau_z(l+1),
 \label{HIsing}
 \eeqa
 where $j_I'= 4J_H |\langle g_1|S_z |g_1\rangle|^2 $. This Hamiltonian may describe the Mn chains on Cu$_2$N/Cu(100) for distant adatoms, a system with $S=5/2$ but with small magnetic anisotropy.

Two obvious  advantages are provided by this approximation. First, the quantum Ising model can be solved analytically in various instances. Second, numerical calculations of the Ising model are significantly faster than those for the complete Heisenberg model. In the two following sections we study the pure decoherence in Ising chains due to the coupling with a d-dimensional electron gas.

\subsection{$T_2$ of  Ising chains  with broken symmetry ground states \label{T2FM}}
We now discuss Kondo induced decoherence for finite size Ising spin chains,  in the regime where symmetry is broken and the system has  two  degenerate ground states, with  finite atomic spin magnetization.   We consider both the ferromagnetic and the antiferromagnetic states, which turn out to be very different. 

We start with the FM case. The chain has  has two ground states with $S_z=\pm N S$.   In the Ising limit,  the chain can be described by Hamiltonian (\ref{HIsing}), and the ground states have all the atomic Ising spins with $\langle \hat \tau_z(l)\rangle= 1$  or $\langle  \hat\tau_z(l)\rangle=-1$. 
Thus, the question that we want to answer is the following. If at $t=0$ the system is prepared in a linear superposition of these two states, how long does it take to decohere?  
Here we assume that temperature is low enough as to make the inelastic contribution negligible, and we compute the adiabatic contribution.

The  resulting  pure decoherence rate $\gamma^{{\rm ad.}}\equiv 1/T_2^*$ can be then written as (see Appendix \ref{appendixd} for details):
\beq
\frac{1}{T_2^* }
 \approx 
\frac{\pi \left(\rho {\cal J}\right)^2}{8\hbar}
k_BT \Lambda^{\rm FM}(k_Fd,N)
\label{T2chainFM}
\eeq
where ${\cal J}=J |\langle g_1|S_z|g_1\rangle|$ and $\Lambda^{\rm FM}(k_Fd,N)$ is a dimensionless  oscillating function of $k_Fd$ that depends on the dimensionality of the electron gas (explicit expressions are given in Appendix~\ref{appendixd}). Figure \ref{fig4} shows $\Lambda^{\rm FM}(k_Fd,N)/N^2$ for one, two and three dimensional fermions. It is immediately apparent that, for small $k_F d$,  the function $\Lambda^{\rm FM}(k_Fd,N)$ tends to  $4N^2$.  In that limit the decoherence rate for the ferromagnetic Ising chain is $N^2$ quicker than the single spin decoherence, given by Eq. (\ref{gadiabatic}).  This can be easily understood realizing that in the limit  $k_Fd=0$, the electrons see the chain as a giant spin with $S_T=N S$, which using Eq. (\ref{chiadb1}),   leads to $\gamma^{{\rm ad.}}\propto N^2S^2$.  Thus, the fragility of coherence scales with the square of the number of atoms in that limit.  
For finite $k_Fd$,  the $N^2$ scaling only survives for $d=1$.  Interestingly, $\Delta^{{\rm FM}}$ is a unique function of $Nk_Fd$ and, for $d=2,d=3$, we finda that  $\gamma^{{\rm ad.}}\propto N^2$ only holds for $Nk_Fd\ll 1$, while in the opposite limit ($Nk_Fd\gg 1$), the decoherence rate from eq. (\ref{T2chainFM})  is linear with $N$.

Let us consider now the decoherence of an antiferromagnetic (AFM) Ising chain described by Hamiltonian (\ref{HIsing0}).
Again, we consider decoherence in the limit where there are two degenerate ground states, the so called N\'eel states.
Thus, we assume a vanishing quantum spin tunneling splitting,  $\Delta_1=0$.  
The pure decoherence rate is given by:
\beq
\frac{1}{T_2^*} 
 \approx 
\frac{\pi \left(\rho {\cal J}\right)^2}{8\hbar}
k_BT \Lambda^{\rm AFM}(k_Fd,N).
\label{T2AFM}
\eeq
The dimensionless functions  $ \Lambda^{\rm AFM}(\zeta,N)$ are plotted in Fig.~\ref{fig4}b)-c) and explicit expressions  are given in Appendix \ref{appendixd}, for $d=1,2,3$. The first thing to notice is that, contrary to the case of the FM chain,  $ \Lambda^{\rm AFM}(k_Fd=0,N)$ is independent of $N$, except for an odd-even effect that can be easily understood. Thus,  the decoherence rate for AFM chains {\em is  not} significantly larger than the one for an individual atom. Actually, the decoherence rate is somewhat smaller. 
 For even (odd)  $N$ the total spin of the chain is zero ($S$).    For even (odd)  $N$, the decoherence rate vanishes (is maximal) in the $k_Fd=0$ limit, which makes sense using the macrospin picture discussed above.
Importantly,  the decoherence rate is an oscillating function of $ N k_F d$.  

These results  can be rationalized as follows.  In the absence of inelastic scattering, complete decoherence between the two N\'eel  states, $|N_1\rangle$ and $|N_2\rangle$,  occurs when  the  wave function of the environment, evolved  interacting with the chain in state $|N_1\rangle$, is orthogonal to the wave function of the environment interacting with the chain in state $|N_2\rangle$~\cite{Stern_Aharonov_pra_1990}.    Since the spin of the quasiparticles is conserved, we can think of the environment as two independent electron gases, with spin $\uparrow$ and $\downarrow$ acting independently.  
For the AFM chain with an even number of atoms,  the effective potential for $\uparrow$ spins is, for both N\'eel states, an alternating potential with attractive and repulsive sites, that just shifts one site when we go from one N\'eel state to the other. 
By contrast, in the case of a FM ground states, the interaction of itinerant quasiparticles with $\uparrow$ spin  feels either a repulsive or an attractive interaction with the chain, depending on which of the two ground states, with $S_z= +N S$ or $S_z= -NS$, is adopted by the chain.

This  provides an intuitive picture as of why the decoherence rate is  much smaller for AFM than for FM chains.   For small $k_Fd$ the conduction electrons are not able to resolve different sites,  let alone  between the 2 N\'eel states, which results in a vanishing decoherence rate.
\begin{figure}[hbt]
\includegraphics[width=1.\linewidth]{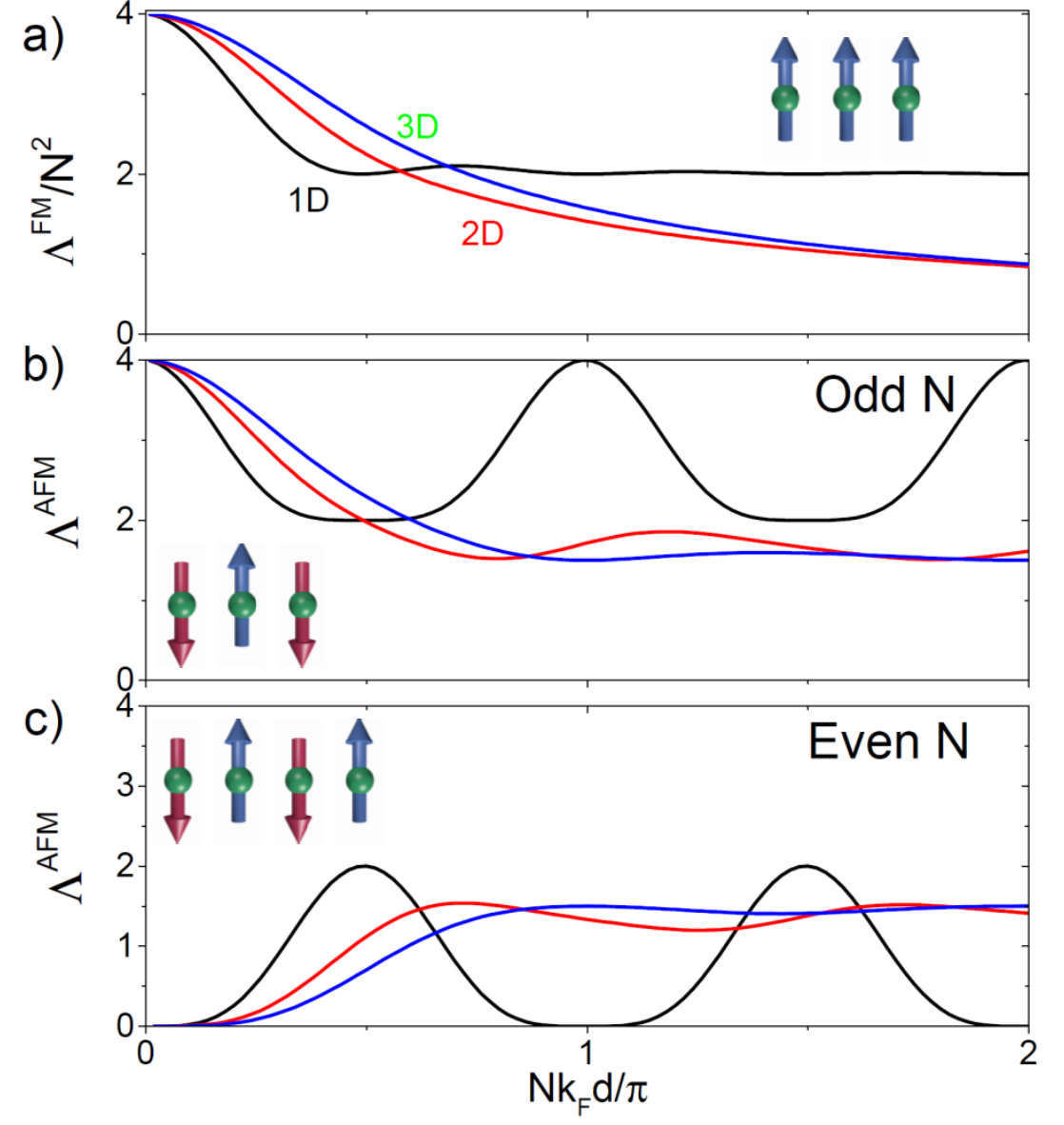}
\caption{ \label{fig4}(Color online) (a) Variation of $\Lambda^{\rm FM}(k_Fd,N)/N^2$ with $Nk_Fd$ for coupling to a one (black line), two (red line) and three (green line)  dimensional electron gas. Notice that for the FM case the dephasing is thus proportional to $N^2$. (b) and (c) Variation of  $\Lambda^{\rm AFM}(k_Fd,N)$ with the parameter $Nk_Fd$ for odd and even spin numbers respectively. The coupling with a one, two  and three dimensional electron gas are denoted as in panel (a). Notice the absence of scaling with $N$ on the vertical axis. In all panels, we have fixed $k_Fd/\pi=0.01$ and changed the chain length $N$.
}
\end{figure}

\section{Discussion and conclusions\label{discuss}}
We have discussed the dissipative spin dynamics of finite size quantum spin chains due to their Kondo coupling to an electron gas. This problem is of paramount importance to understand a variety of outstanding experiments where spin chains can be crafted, atom by atom,    choosing (with some limitations)  the quantum spin of each site $S$, the number of atoms in the chain $N$, the sign of the  spin-spin Heisenberg interactions $J_H$ , and the spin-spin distance $d$. 

We have addressed the problem within the standard  Bloch Redfield theory,  so that Kondo interactions are considered up to second order  in $J$.  At this level, there are    two well established important results in condensed matter physics. First,  the so called Korringa  mechanism for spin relaxation, that states that the spin relaxation rate $T_1^{-1}$ for a  single localized spin because of its Kondo coupling $J$ to itinerant electrons with density of states at the Fermi energy $\rho$ is proportional to $(\rho J)^2$.  Second, the so  called RKKY exchange interaction that arises when two otherwise decoupled local spins are exchanged coupled to the same electron gas. The strength of this interaction is also proportional to $(\rho J)^2$ but its magnitude is an oscillating function of $k_F d$, where $k_F$ is the Fermi wavenumber and $d$ is the distance between the spins. 

The main result of this work is  the finding  that  the spin relaxation and decoherence of spin chains  has an oscillatory dependence on 
$k_F d$, ie, it has  RKKY-type oscillations.  Given that both $T_1$ and $T_2$ play a central role to determine the use of these designer nanomagnets as magnetometers, as shown in recent experiments~\cite{Baumann_Paul_science_2015,Yan_Malavolti_arxiv_2016,natterer2016},  our results suggest that it might be possible to engineer these quantities by controlling either $k_F$, via gating using graphene as a substrate, or changing $d$.  In addition, we have also discussed how $T_1$ and $T_2$ depend on the number of spin $N$ in a chain,  which provides yet another control parameter for quantum engineering.

The second important result of this paper  is 
the deep underlying origin of the RKKY-like oscillations of $T_1$ and $T_2$ in spin chains: within the Bloch-Redfield theory, the dissipative coupling to an environment produces both a dissipative response, that yields finite $T_1$ and $T_2$, and also a reactive response that produces a shift of the energy levels.
In the context of quantum electrodynamics,  
the  coupling to photons results in the dissipative  decay of the optically induced coherence, ie,  a finite   $T_1$ and $T_2$ times, 
together with a shift  of the energy levels, the   Lamb shift. 
  In the context of individual magnetic atoms,  Kondo interactions yield a dissipative effect,   Korringa spin relaxation, and less frequently discussed reactive response,  the renormalization of the $g$ factor~\cite{Delgado_Hirjibehedin_sci_2014} and, in the case of anisotropic atoms, the  recently observed renormalization of the magnetic anisotropy~\cite{Oberg_Calvo_natnano_2013,Jacobson_Herden_natcom_2015}.  In this paper we extend this unifying picture  on two counts. First, we note  that  the RKKY interaction can be understood as the Lamb shift of the energy levels of a spin chain, or for that matter, any other magnetic nanostructure described by Eq. (\ref{shiftRKKY}). Second,  we realize that the dissipative counterpart of the RKKY coupling, the spin relaxation and decoherence  rates, $T_1^{-1}$ and $T_2^{-1}$ are also oscillating functions of $k_F d$.  
The oscillatory dependence of the spin relaxation on $k_F d$ can also be seen from the perspective of the itinerant quasiparticles that drive it.   
Both $T_1$ and $T_2$ reflect how often itinerant quasiparticles are scattered by the spin chain.   
Thus, engineering the quasiparticle wave functions on the  surface can result in significant changes of the dissipative atomic spin chain dynamics.

The notion that the spin relaxation of a spin nanostructure presents RKKY oscillations may be anticipated using an alternative argument. By using perturbation and linear response theories,  
 the spin relaxation of a local moment located at $\vec{r}_i$,  which could be an electronic or nuclear spin,  induced by its exchange coupling to itinerant electrons,
  can be written down in terms of the local spin susceptibility of the itinerant electrons, $\chi_{\rm el}(\vec{r},\vec{r}_i)$~\cite{Moriya1956,Moriya1963}.  
     Extension of this perturbative calculation to the case of two or more local impurities at sites $\vec{r}_1$, $\vec{r}_2$,   naturally leads to a expression of the rate that includes both, diagonal terms in the susceptibility matrix as well as the non-diagonal entry  $\chi_{\rm el}(\vec{r}_1,\vec{r}_2)$.  For a free electron gas, it can be easily seen that the non-diagonal spin susceptibility is an oscillating function of the relative coordinate $|\vec{r}_1-\vec{r}_2|$. Thus, the results of our paper may be connected, to some extent, with previous works computing 
    Gilbert damping in ferromagnetic multilayers  using the concept of dynamical RKKY, between two distant ferromagnetic layers~\cite{Simanek_Heinrich_prb_2003}.

 Our results  are expected to apply for more complicated geometries, such as ladders and rings, and probably to physical phenomena not captured by the second order treatment of the Kondo coupling $J$.  Thus,   the engineering of the Kondo effect with elliptical corrals~\cite{manoharan2000} could also work to engineer spin relaxation of individual magnetic atoms and spin arrays as well.  This whole picture requires a proper description of the phase factors $e^{i(\vec{k}-\vec{k}')\cdot\vec{r}}$ in the Kondo interaction with a  multi-spin structure, as we do here, and it might help to account for the  peculiar experimental observations of spatial modulations in the Kondo peak of MnFe chains with an odd number of Mn and one Fe atom at the edge~\cite{Choi_Robles_arXiv_2015}. 

 Our study highlights the very different dissipative  properties of different quantum spin states on the same structure, illustrated for instance with the spin dimer.  This connects with the concept of subradiant and superradiant states for ensembles of atoms coupled to photons.    Moreover, we show how different confined spin waves in short ferromagnetic chains~\cite{Spinelli_Bryant_naturematerials_2014} have a very different spin relaxation rates. Specifically, for small $k_F d$, the  decay rate of the first spin wave should be a linearly growing function of the chain size, similar to the superradiant regime in quantum optics. By contrast, the decay of the second spin wave excitation, which is spatially modulated, is quenched due to the destructive interference of the scattering with neighboring spins, analogue with the subradiant regime.  

Our work establish a connection between the dissipative  dynamics of quantum nanomagnets and QED systems:  spins play the role of atomic levels,  electron-hole excitations across the Fermi level play the role of photons, and Kondo exchange plays the role of electron-photon interactions. 
 With all these ingredients, we provide the principles of design for future quantum technologies based on magnetic nanostructures, such as quantum magnetometers.

\acknowledgments
JFR acknowledges financial supported by MEC-Spain
(FIS2013-47328-C2-2-P) and Generalitat Valenciana
(ACOMP/2010/070), Prometeo.
This work is funded by ERDF funds through the Portuguese Operational Program for Competitiveness and InternationalizationÐ COMPETE 2020, and National Funds through FCT- The Portuguese Foundation for Science and Technology, under the project "PTDC/FIS-NAN/4662/2014" (016656).
FD  acknowledges support from  Spanish Government
through grants FIS2013-473228 and MAT2015-66888-C3-2-R. We thank N. Lorente, S. Loth  for fruitful conversations, during SPICE  workshop on ``Magnetic adatoms as building blocks for quantum magnetism". FD acknowledges hospitality of the Departamento de F\'{i}sica Aplicada, Universidad de Alicante.

\appendix

\section{Origin of the phase in the Kondo Hamiltonian\label{appendixA}}
 The Kondo coupling can arise from two sources, direct and kinetic exchange. The direct  $sd$ exchange arises whenever states $s$ and $d$  overlap in space~\cite{Kondo_ptp_1964}. 
 The resulting exchange Hamiltonian  is proportional to the itinerant spin density $\vec s(\vec{r}_l)$ evaluated at the atomic spin center, given the short-range nature of exchange, and it is given by
\beqa
\vec s(\vec{r}) = \sum_{\lambda,\lambda',\sigma,\sigma'} 
\phi_{\lambda}^*(\vec{r}) \phi_{\lambda'}(\vec{r})
\frac{\vec{\sigma}_{\sigma\sigma'}}{2}
 c^{\dagger}_{\lambda,\sigma} c_{\lambda'\sigma'}, 
\label{SDEN}
\eeqa
where $\phi_{\lambda}(\vec{r})$ are the single particle states associated to $\lambda$, which we take as the wave vector $\vec{k}$ in the rest of the paper. 
For itinerant electrons in a crystal, for which momentum $\vec{k}$ is a good quantum number,  we have
$\phi_{\lambda}(\vec{r})\propto e^{i\vec{k}_{\lambda}\cdot\vec{r}}$.
 We thereby ignore band indexes and we also put aside the Kondo interactions with the STM tip.  By so doing, we give up the possibility of describing tip-induced relaxation as well as spin-flip assisted transport between tip and surface, which could be treated on a similar footing.

In the case of the kinetic exchange~\cite{Anderson_prl_1966,Appelbaum_prl_1966}, the Kondo interaction is proportional to the square of the hybridization $V_{\vec k}$ between the localized $d$-electrons and the itinerant electrons wavefunction  with wavevector $\vec{k}$~\cite{Schrieffer_Wolff_pr_1966}.
The hybridization matrix element picks a Bloch phase between different atoms, $V_{\vec{k}}(\vec{r}_l)= V_{\vec{k}}(\vec{r}_{l'})e^{\vec{k}\cdot (\vec{r}_l-\vec{r}_{l'})}$.   When the  Kondo coupling is derived from the Anderson model by means of a canonical transformation~\cite{Schrieffer_Wolff_pr_1966},  the  Kondo interaction $J(k,k')$ is proportional to $V_{\vec{k}}^* V_{\vec{k}'} $. Thus, the Kondo interaction will takes the general form of Eq. (\ref{VKONDO}).

\section{Explicit expressions of the Bloch-Redfield tensor\label{appendixB} }

In the case of a spin array with Hamiltonian ${\cal H}_{\rm chain}$ coupled to a free electron gas through the  
Kondo interaction (\ref{VKONDO}), writing the Bloch-Redfield tensor components ${\cal R}_{NMKL}$ as the sum $ {\cal R}_{NMKL}^+ + {\cal R}_{NMKL}^-$, one finds that~\cite{Delgado_Rossier_rev}:
\beqa
{\cal R}_{NMKL}^\pm = \frac{1}{\hbar^2}\sum_{\vec k\vec k'}f(\epsilon_{\vec k})\left(1-f(\epsilon_{\vec k'})\right)\Sigma_{NMKL}^\pm(\vec k,\vec k')
\crcr
\label{RpmK}
\eeqa
where, using the short notation $\omega_{\vec k\vec k'}=(\epsilon_{\vec k}-\epsilon_{\vec k'} )/\hbar$ for the energy difference of the single particle wavefunctions in the electronic bath, we have
\beqa
\Sigma_{NMKL}^+(\vec k,\vec k')&=&-2i \sum_{a,nn'}J(n)J(n'){\cal F}_{n-n'}(\vec k-\vec k')
\crcr
&&\hspace{-2.5cm}\times
\left[
\frac{ S^a_{LM}(n)S^a_{NK}(n')}{\omega_{NK}-\omega_{\vec k\vec k'} - i0^{+}}
- \delta_{LM}\sum_R 
 \frac{S^a_{NR}(n) S^a_{RK}(n')}
{\omega_{RK}-\omega_{\vec k\vec  k'} - i0^{+}}
\right].
\crcr
&&
\label{sigmap}
\eeqa
Here we have introduced the functions ${\cal F}_n(\vec q)= e^{i\vec q\cdot \vec r_n}$, while $i0^+$ denotes an infinitesimal (positive) imaginary number.
 Similarly, one has
\beqa
\Sigma_{NMKL}^-(\vec k,\vec k')&=&-2i\sum_{a,nn'}J(n)J(n'){\cal F}_{n-n'}^*(\vec k-\vec k')
\crcr
&&\hspace{-2.5cm}\times
\left[ \frac{S^a_{LM}(n) S^a_{NK}(n')}
{ \omega_{ML}+\omega_{\vec k\vec k'}  - i0^{+}}
- \delta_{NK}\sum_R
 \frac{S^a_{LR}(n)S^a_{RM}(n') }{\omega_{RL}+\omega_{\vec k\vec k'}- i0^{+}}
\right].
\crcr
&&
\label{sigmam}
\eeqa
The tensor components ${\cal R}_{NMKL}$ has both a real and an imaginary part. They can be easily split by taking into account that
\beq
\frac{-i}{\epsilon-\epsilon'-i0^+}=\pi\delta(\epsilon-\epsilon')-i{\cal P}\frac{1}{\epsilon-\epsilon'},
\label{Pv}
\eeq
where ${\cal P}$ stands for the Cauchy principal value. Hence, the real part will be the responsible of the transition and decoherence rates, Eqs. (\ref{dinteg}) and (\ref{invT2chain}), and it recovers the result of the Fermi Golden Rule. On the other hand, the imaginary part, which is associated to the principal part of the integral over frequencies, is associated to the energy shifts.

The only possible non-zero contributions to the energy shifts are given by the components of the form ${\cal R}_{NMNM}$, whose elements $\Sigma_{NMNM}(\vec k,\vec k')$ satisfy
\begin{widetext}
\beqa
{\rm Im}\left[\Sigma_{NMNM}^+(\vec k,\vec k')+\Sigma_{NMNM}^-(\vec k,\vec k')
\right]
&=&-{\cal P}\frac{2}{\omega_{\vec k\vec k'}}   \sum_{a,nn'}J(n)J(n') 
S^a_{MM}(n)S^a_{NN}(n')  
2 i\sin\left[ (\vec k-\vec k')\cdot (\vec r_n-\vec r_{n'})\right]
\crcr
&&\hspace{-4.5cm}+ 2{\cal P}\sum_{a,nn'} J(n)J(n') \sum_R
\left[  {\cal F}_{n-n'}(\vec k-\vec k') \frac{ S^a_{NR}(n) S^a_{RN}(n')    } {\omega_{RN}-\omega_{\vec k\vec  k'}}
- {\cal F}_{n-n'}^*(\vec k-\vec k')  \frac{S^a_{MR}(n)S^a_{RM}(n') }
{\omega_{RM}-\omega_{\vec k\vec k'}}
\right].
\label{sigmapS}
\eeqa
\end{widetext}
The first term in Eq. (\ref{sigmapS}) identically cancels when doing the sum over the positions $n$ and $n'$ since it is an odd function of $n-n'$. Thus, one can write down the  tensor components in the form (\ref{2sides}),  with the energy shift $\delta\omega_M$  experienced by an state $|M\rangle$ given by
 \beqa
 \delta \omega _{M}&=&\frac{1}{\hbar^2}{\cal P}\sum_{\vec k\vec k'}f(\epsilon_{\vec k})\left(1-f(\epsilon_{\vec k'})\right)\sum_{R}\frac{1}{\omega_{\vec k\vec k'}+\omega_{MR}}
\crcr
&&\hspace{-0.8cm}\times
2\sum_{a} 
\left| \sum_n J(n) S^a_{MR}(n)e^{i(\vec k-\vec k')\cdot \vec r_n}
\right|^2.
 \eeqa

\subsection{Dimensionality of the electron gas}
The relaxation and decoherence rates involves integration over the Fermi surface since the product
 $f(\epsilon_{\vec k})\left(1-f(\epsilon_{\vec k'})\right)$ is non-zero only in the vicinity of the Fermi level.
This implies that one can approximate the wavevectors $\vec k$ and $\vec k'$ in Eq. (\ref{RpmK}) by its value on the Fermi surface, i.e., $\vec k\approx k_F \hat k$ and $\vec k'\approx k_F \hat k'$, where $\hat k=\vec k/|\vec k|$. 
Thus, it is convenient to define the average of $\chi_{M,M'}(\vec q)$ over the Fermi surface,
\beq
\chi_{M,M'}^ {k_F}=\frac{1}{\Omega_ d^2}\int d \hat k d \hat k'\chi_{M,M'}\left(k_F(\hat k-\hat k')\right),
\label{LambdaG}
\eeq
where $\Omega_d=\int d\hat k$. (Notice that in one dimension, this is nothing else that the sum over the forward and backwards components).

One can get simple analytical expression for the angular integration in the case of linear spins chains, as illustrated in the dimer case, Eqs. (\ref{gammastd}-\ref{gammattd}).
These expressions will depend on the dimensionality of the electron gas.

\section{$T_2$ in Ising chains\label{appendixd}}
The expressions of $\gamma_{M,M'}^{\rm ad.}$ are much simpler in the case of Ising chains for which $S^ \pm_{MM'}(l)=0$, and only the $S^z_{MM'}$ components of the spin gives a non-zero contribution.
Thus, one can write
\beqa
\chi_{M,M'}^ {ad.}(\vec{q})&\equiv& 
\left|\sum_{l}\left(e^{i\vec{q}\cdot \vec r_l}  S_{MM}^z(l)
- e^{-i\vec{q}\cdot \vec r_l}  S_{M'M'}^z(l)\right)\right|^2.
\nonumber
\eeqa

\subsection{$T_2$ in FM Ising chains}
Let us assume that states $M,M'$ are such that $S^ z_{MM}(l)=+S$ and $S^ z_{M'M'}(l)=-S$. Thus, one gets that 
\beqa
\chi_{MM'}^ {ad.}(\vec q)=4S^2\sum_{ll'}\cos \left(\vec q\cdot \vec r_l\right)\cos \left(\vec q\cdot \vec r_{l'}\right)
\eeqa
In order to extract the adiabatic decoherence rate $1/T_2^*$, it is convenient to start by the average over the Fermi surface and then, if possible, making the explicit sum over sites $l,l'$.  First, we write the pure decoherence rate in the form
\beqa
\gamma^{\rm ad.}_{MM'}=\frac{\pi(\rho {\cal J})^2}{8\hbar}k_BT \Lambda_{MM'}^{\rm FM}(k_Fd,N),
\eeqa
where ${\cal J}=JS$.
For coupling to a one dimensional electron gas one gets fully analytical results
\beqa
\Lambda_{1d}^ {\rm FM} (\xi,N)&=&2 \Big[N^2
\crcr
&&\hspace{-1.5 cm} +\csc ^2(\xi) \sin
   ^2(\xi N) \cos ^2(\xi
   (N+1))\Big].
\label{lambdafm}
\eeqa
This expressions has two very interesting limits when $N\gg 1$. For $k_Fd=n\pi, n=0,1,\dots$, its leads to $ \Lambda_{1d}^ {\rm FM} (k_Fd)\sim 4N^2$, while for $|k_Fd-n\pi|\ge \pi/N$ with $n$ integer, $ \Lambda_{1d}^ {\rm FM} (k_Fd)\sim 2N^2$.

For two and three dimensions, the expressions are left in term of explicit sums over the adatoms positions:
\beqa
\Lambda_{2d}^ {\rm FM} (\xi,N)&=&2 \sum_{ll'}\left[J_0^2\left(\xi|l-l'|\right)
+J_0^2\left(\xi|l-l'|\right)\right],
\nonumber
\eeqa
and 
\beq
\Lambda_{3d}^ {\rm FM} (\xi,N)=2\sum_{ll'}\left[{\rm Sinc}^2\left(\xi|l-l'|\right)
+{\rm Sinc}^2\left(\xi|l-l'|\right)\right].
\eeq

\subsection{$T_2$ in AFM Ising chains}
Let us assume that states $M,M'$ are the classical N\'eel states with $S^ z_{MM}(l)=(-1)^ lS$ and $S^ z_{M'M'}(l)=(-1)^ {l+1}S$. For an even number $N$, one has
\beqa
\Lambda_{1d}^ {\rm AFM} (\xi,N)&=&\frac{1}{2}{\rm csc}^2(2\xi)\Big(
\sin(2\xi)+\sin(2N\xi)
\crcr
&-&\sin(2(1+N)\xi)\Big)^2,
\eeqa
while for odd-chains
\beqa
\Lambda_{1d}^ {\rm AFM} (\xi,N)&=&\frac{1}{2}\Big[4+
\big\{1+{\rm csc}^2(2\xi)\left(-\sin(2N\xi)
\right.
\crcr
&-& \left. \sin(2(1+N)\xi)\right)\big\} \Big].
\eeqa
Interestingly, in the case of even-number chains, one obtain the following relation
\beqa
\Lambda_{1d}^ {\rm AFM} (\xi,N)&=&\Lambda_{1d}^ {\rm FM} (\xi+\pi/2)-2N^ 2.
 \label{lambdaafm}
\eeqa
As in the FM case, $\Lambda_{1d}^ {\rm AFM} (\xi,N)$ is an periodic function of $k_Fd$ with period $\pi$, with the notable difference that its maxima occurs at $k_Fd=\pi/2$, where $ \Lambda_{1d}^ {\rm AFM} (\pi/2,N)=2N^2$. By contrast, 
around $k_Fd=0$, we have that  
 $\Lambda_{1d}^ {\rm AFM} (k_Fd,N)\approx 0$ for $N$ even and $\Lambda_{1d}^ {\rm AFM} (k_Fd,N)\approx 4$ for odd $N$.

In two and three dimensions the expressions are left in term of explicit sums over the adatoms positions:
\beqa
\Lambda_{2d}^ {\rm AFM} (\xi)&=&2 \sum_{ll'}(-1)^{l+l'}\Big[J_0^2\left(\xi|l-l'|\right)
\crcr
&&+J_0^2\left(\xi|l-l'|\right)\Big],
\nonumber
\eeqa
and 
\beqa
\Lambda_{3d}^ {\rm AFM} (\xi)&=&2 \sum_{ll'}(-1)^{l+l'}\Big[{\rm Sinc}^2\left(\xi|l-l'|\right)
\crcr
&+&{\rm Sinc}^2\left(\xi|l-l'|\right)\Big].
\nonumber
\eeqa





\end{document}